\documentclass[11pt]{article}

\usepackage{pdflscape} 
\usepackage{amsthm}
\usepackage{floatrow}
\usepackage[margin=1in]{geometry}
\usepackage[comma]{natbib} 
\usepackage{anysize}
\usepackage[usenames, dvipsnames]{color}
\usepackage{xcolor,color}
\usepackage{graphicx}
\usepackage{amsmath,amssymb,amsthm,amsopn,mathrsfs}
\usepackage[pdftex,bookmarks,colorlinks]{hyperref}
\usepackage{caption}
\usepackage{subcaption}
\usepackage[applemac]{inputenc}
\usepackage{booktabs}
\usepackage[makeroom]{cancel}
\usepackage{dsfont}
\usepackage{accents}
\usepackage{xcolor,color}
\usepackage{epstopdf}
\usepackage{multirow}
\usepackage{enumitem}
\usepackage{adjustbox}

\graphicspath{{./figures/}}

\definecolor{navy}{rgb}{0,0,0.502}

\hypersetup{colorlinks,%
citecolor=navy,%
filecolor=navy,%
linkcolor=navy,%
urlcolor=navy,%
}

\newcommand{\bX}{{\boldsymbol{X}}}
\newcommand{\bx}{{\boldsymbol{x}}}
\newcommand{\bb}{{\boldsymbol{b}}}
\newcommand{\bS}{{\boldsymbol{S}}}
\newcommand{\bB}{{\boldsymbol{B}}}
\newcommand{\bs}{{\boldsymbol{s}}}
\newcommand{\bz}{{\boldsymbol{z}}}
\newcommand{\bi}{{\boldsymbol{i}}}
\def\real{{\mathbb R}}

\theoremstyle{definition}
\theoremstyle{plain}
\theoremstyle{remark}

\newtheorem{mydef}{Proposition}
\theoremstyle{definition}

\begin{document}

\title{Composite likelihood methods for histogram-valued\\ random variables}

\author{T. Whitaker\footnote{School of Mathematics and Statistics, University of New South Wales, Sydney.},\; B. Beranger$^{*}$ and S. A. Sisson$^{*}$}
\date{}

\maketitle
\begin{abstract}
Symbolic data analysis has been proposed as a technique for summarising large and complex datasets into a much smaller and tractable number of distributions -- such as random rectangles or histograms -- each describing a portion of the larger dataset. Recent work has developed likelihood-based methods that permit fitting models for the underlying data while only observing the distributional summaries. 
However, while powerful, when working with random histograms this approach  rapidly becomes computationally intractable as the dimension of the underlying data increases.
We introduce a composite-likelihood variation of this likelihood-based approach for the analysis of random histograms in $K$ dimensions, through the construction of lower-dimensional marginal histograms.
The performance of this approach is examined through simulated and real data analysis of max-stable models for spatial extremes using millions of observed datapoints in more than $K=100$ dimensions. Large  computational savings are available compared to existing model fitting approaches.
\\

\\

%
%
%
\end{abstract}

\section{Introduction}
\label{intro}

Continuing advances in measurement technology and information storage are leading to the creation of increasingly large and complex datasets. This inevitably brings new inferential challenges.
Symbolic data analysis (SDA), a relatively new field in statistics, has been developed as one way of addressing these issues \citep[e.g.][]{diday1989, bock2000}. In essence, SDA argues that many important questions can be answered without needing to observe data at the micro-level, and that higher-level, group-based information may be sufficient. As a result, SDA methodology aggregates the micro-data into a much smaller number of distributional summaries, such as random rectangles, random histograms and categorical multi-valued variables, each summarising a portion of the larger dataset \citep{dias+b15,rademacher2017,billard+d06}. These new data ``points'' (i.e.~distributions) are then analysed directly, without any further reference to the micro-data. 
See e.g.~\cite{billard2011}, \cite{bertrand2000} and \cite{billard+d03} for an exposition of these ideas.

SDA methods have found wide application in current statistical practise, and have been developed for a range of inferential procedures, including regression models \citep{dias+b15}, principle component analysis  \citep{kosmelj2014}, time series analysis \citep{wang+zl16}, clustering \citep{brito2015probabilistic}, discriminant analysis \citep{silva+b15}, Bayesian hierarchical modelling \citep{lin+cs17} \textcolor{black}{and logistic regression \citep{whitaker+bs19}}. Likelihood-based methods for distributional data were introduced by \cite{lerademacher+b10} and \cite{brito+d12} for direct modelling at the level of the distributional summary.

More recently, \cite{zhang+bs16} and \cite{beranger+ls18} developed likelihood functions for observed random rectangles and histograms that directly accounts for the process of constructing the symbols from the underlying micro-data. By explicitly considering the full generative process -- from micro-data generation to constructing the resulting distributional summary -- the resulting symbolic likelihood allows the fitting of the standard micro-data likelihood, but while only observing the distributional-based data summaries. The symbolic likelihood reduces to the standard micro-data likelihood as the observed symbols reduce to the underlying micro-data (e.g.~as the number of histogram bins gets large, and the size of each histogram bin gets small). \cite{beranger+ls18} demonstrate a $14\times$ computational speed up for the symbolic analysis over the standard micro-data analysis for computing the maximum likelihood estimates of a  hierarchical skew-normal model.
\textcolor{black}{The approach of \cite{zhang+bs16} and \cite{beranger+ls18}, and the one taken here, differs from the standard usage of SDA in that it is primarily focused on fitting models for the micro-data, given the observed symbols, rather than on fitting models for the symbols themselves.}

While attractive, a limitation of this approach is that grid-based multivariate histograms become highly inefficient as data summaries as the dimension of the data increases. This means that the histogram-based approach in \cite{beranger+ls18}, where the  computational overhead is proportional to the number and dimension of histogram bins, is practically limited to lower-dimensional data analyses.

In this paper we address this problem by extending the likelihood-based approach of  \cite{beranger+ls18} to the composite-likelihood setting. Focusing on histogram-based distributional summaries, the components of the composite likelihood are constructed based on low-dimensional marginal histograms derived from the full $K$-dimensional histogram.
We demonstrate consistency of the resulting symbolic composite maximum likelihood estimator, and 
show that for a certain level of data aggregation, the symbolic composite likelihood function provides a useful and more computationally efficient substitute for the standard micro-data analysis. 
We obtain results that describe the reduction in information that occurs when aggregating the micro-data into histograms, and how this reduction is dependent on the number of observed histograms.
These results also provide insights on the efficiency of standard composite likelihood techniques when the micro-data are grouped into blocks, but where the location of data within each block is not known.

While the above techniques are general, throughout we are motivated by the need to develop computationally viable statistical techniques for fitting max-stable process models for spatial extremes. This becomes particularly challenging when both the number of spatial dimensions $K$ (the number of physical recording stations) and the number of observations ($N$) become large, as is the case with millennial scale climate simulations \citep{Huang2016}. While composite-likelihood techniques \citep{padoan+rs10,Blanchet2011,varin+rf11,Lee2013,cast2016,beranger+ss19} provide one way to approach the issue of spatial dimensions, they are not able to cope with large amounts of observed data at each spatial location. By developing composite likelihood techniques for the analysis of $K$-dimensional histogram-valued random variables, we are able to directly and efficiently fit max-stable processe models to very large temporal datasets.

This article is structured as follows: 
In Section \ref{sec:2} we describe the ideas behind the symbolic likelihood framework of \cite{beranger+ls18}, with a focus on histogram-valued random variables,  extend this approach to the case of a marginal histogram, and briefly present relevant background on composite likelihood methods.

In Section \ref{sec:compsym} we extend the histogram-based  symbolic likelihood function to the composite likelihood setting. We demonstrate that increasing the number of bins (and reducing their size) in each histogram yields {\color{black}composite maximum likelihood estimators (MLEs)} that are asymptotically consistent with those of the classical (micro-data) setting, but at a potentially much cheaper computational cost. While these {\color{black}composite MLEs} retain this asymptotic consistency regardless of the method of histogram construction (as long as the volume of each bin approaches zero as the number of bins approaches infinity) and how many random histograms are used, their variances depend heavily on the amount of temporal information retained during the data aggregation process. Accordingly we show that increasing the number of random histograms leads to an overall decrease in the variance of the composite MLE. 
In Section \ref{sec:examples} we explore the performance of the histogram-based composite likelihood function through simulation studies using max-stable processes, and in Section \ref{sec:6} we analyse
real and future-simulated datasets comprising daily maxima temperature data from $105$ locations across Australia.
We conclude with a Discussion.

\section{Symbolic and composite likelihoods}
\label{sec:2}

We first provide a brief overview of likelihood-based methods for {\em symbolic} random variables, in particular focusing on histogram-valued random variables and the approach of \cite{beranger+ls18}. Motivated by a desire to reduce computational overheads as the dimension of the histogram $K$ increases, we extend this setup to the case of a  {\em marginal}-histogram (i.e.~a lower-dimensional margin of an original histogram). We then briefly review the ideas behind composite likelihoods in a general setting.

\subsection{Generative symbolic likelihoods}
\label{symbolicmodel} 

In simple terms, symbolic random variables are distributional-valued random variables that are constructed by the aggregation of standard, classical random variables into a distributional summary form, such as a random interval or random histogram. Symbolic data analysis is the study and analysis of symbolic random variables \citep{billard2011,billard+d03,bock2000}.
Within this field, two main likelihood-based techniques have been developed for the analysis of symbolic data; one based on analysing the symbols directly \citep{lerademacher+b10,brito+d12,lin+cs17} and one based on also modelling the construction of the symbols from the generating process of the classical random variables \citep{beranger+ls18,zhang+bs16}.
This latter technique allows for the use of symbolic data analysis methods as a means to expedite standard data analyses for large and complex datasets. We adopt both this approach and motivation here.

The general construction of \cite{beranger+ls18} is given as follows.
Denote by $\bX=(X_1,\ldots,X_N)$ a vector of i.i.d.~classical random variables, which takes values in some space $D_\bX$ and has density $g_\bX(\,\cdot\,;\theta)$ with unknown parameter vector $\theta$. Each $X_i$ takes values in $D_X$ and has density $g_X(\,\cdot\,;\theta)=\int g_\bX(\,\cdot\,;\theta)d\bX_{-i}$ where $\bX_{-i}=\bX/X_i$, so that $D_\bX=(D_X)^N$. The observed values $\bx$ of $\bX$ can then be aggregated into a distribution-valued symbol $s$, itself a realisation of some symbolic random variable $S \in D_S$, according to a known function $f_{S|\bX=\bx}(s|\bx, \phi)$. The likelihood associated with the process of generating and constructing the observed symbol $s$ is then given by
\begin{equation}
L(s;\theta, \phi) \propto \int_{D_{\bX}}f_{S|\bX=\bx}(s|\bx, \phi)g_{\bX}(\bx;\theta)d\bx.
\label{eq:1}
\end{equation}
That is, $L(s ; \theta,\phi)$ is the expectation of the classical data likelihood $g_{\bX}(\bx;\theta)$ over all possible classical datasets $\bx$ that could have produced the observed symbol $s$.

\cite{beranger+ls18} considered several forms for $f_{S|\bX=\bx}(s|\bx, \phi)$ that allowed for different types of symbol (e.g.~random intervals, hyper-rectangles and different forms of random histogram) and accordingly different resulting forms of symbolic likelihood function.
Here we focus on the fixed-bin, random-counts histogram, although extension of the results in this article to other symbolic likelihood forms is possible.

Suppose that $X_1,\ldots,X_N$ are $K$-dimensional random vectors with
$D_X = \mathbb R^K$. The collection of $N$ classical data observations $\bx \in \mathbb R^{N\times K}$ may be aggregated into a $K$-dimensional histogram on $D_X$, where the $k$-th margin of $D_X$ is partitioned into $B^k\in\mathbb{N}$ bins, so that $B^1\times \cdots \times B^K$ bins are created in $D_X$ through the $K$-dimensional intersections of each marginal bin.
Indexing each bin $\bb = (b_1,\ldots,b_K)$, $b_k = 1,\ldots,B^k$, as the vector of marginal bin indices,
 bin $\bb$ may be constructed over the space 
$\Upsilon_{\bb} =  \Upsilon_{\bb}^{1}\times \cdots \times \Upsilon_{\bb}^{K}$, where $\Upsilon_{\bb}^k = (y_{b_{k}-1}^k,y_{b_k}^k]\subset\mathbb{R}$, and where, for each margin $k$,
$-\infty<y^k_0<y^k_1<\ldots<y^k_{B^k}<\infty$ are fixed points that define the change from one bin to the next.
That is, $\bb$ describes the coordinates of a bin within the $K$-dimensional histogram and  $\Upsilon_{\bb}\subseteq{\mathbb R}^K$ defines the space that it covers. 

Now let $S_\bb$ denote the random number of observed data points $X_1,\ldots,X_N$ that fall in bin $\bb$.
Then $\bS=(S_{\bold 1},\ldots,S_{\bB})$ is the vector of counts from the first bin $\bold 1 = (1,\ldots,1)$ to the last bin $\bB = (B^1,\ldots,B^K)$, of length $B^1\times \cdots\times B^K$, and which satisfies $\sum_{\bb}S_{\bb}=N$. That is, $\bS$ is a random histogram with $N$ observations. Following \cite{beranger+ls18}, and assuming that $g_{\bX}(\bx;\theta)=\prod_{i=1}^Ng_X(x_i;\theta)$, the resulting symbolic likelihood function \eqref{eq:1} then becomes
\begin{equation}
\label{histlik}
L(\bs;\theta) \propto \frac{N!}{s_{\bold 1}!\ldots s_{\bB}!}\prod_{\bb=\bold 1}^{\bB}P_{\bb}(\theta)^{s_{\bb}},
\end{equation}
where $\bs=(s_{\bold 1},\ldots,s_\bB)$ is the observed value of $\bS$, and
where $P_{\bb}(\theta) = \int_{\Upsilon_{\bb}}g_\bX(\bz;\theta)d\bz$ is the probability of observing a datapoint in bin $\Upsilon_\bb$ under the model $g_\bX(\bx;\theta)$. (The $\phi$ parameter in \eqref{eq:1}, which controls quantities relevant to constructing the symbol,  is fixed in this setting, and so we omit it from subsequent notation.)
This multinomial form of likelihood makes intuitive sense in that maximising this likelihood amounts to choosing parameters $\theta$ that optimally match the empirical bin proportions with the corresponding bin probabilities under the model $g_\bX(\bx;\theta)$.

Looking ahead to Section \ref{sec:compsym} where we will be constructing composite symbolic likelihood functions, 
suppose that
we are only interested in a subset of the $K$ dimensions, represented by some index set $\bi = (i_1,\ldots,i_I)\subseteq \{1,\ldots,K\}$, where for convenience $i_1<\ldots<i_I$. 
We may then construct the associated $I$-dimensional marginal histogram, defining
$\bb^{\bi}$ as the subvector of $\bb$ containing those elements corresponding to the index set $\bi$.
(We use this notation more generally, so that a vector with superscript $\bi$  means the subvector containing those elements corresponding to the index set $\bi$.)
Then if $S_{\bb^\bi}^\bi$ is the random number of observed data points $X_1^\bi,\ldots,X_N^\bi$ that fall in bin $\bb^\bi$, 
we may construct an $I$-dimensional random {\em marginal} histogram $\bS^\bi=(S_{{\bold 1}^\bi}^\bi,\ldots,S_{\bB^\bi}^\bi)$  as the associated vector of random counts from the first bin
${\bold 1}^\bi=(1,\ldots,1)$ to the last bin $\bB^{\bi} = (B^{i_1},\ldots,B^{i_I})$.
The vector $\bS^\bi$ has length $B^{i_1}\times\ldots\times B^{i_I}$ and satisfies $\sum_{\bb^\bi}S_{\bb^\bi}^\bi=N$.

Note that we can write $\bS^\bi_{\bb^\bi}=\sum_{\tilde{\bb}:\tilde{\bb}^\bi=\bb^\bi}\bS_{\tilde{\bb}}$ so that we are effectively marginalising out the non-indexed set $-\bi=\{1,\ldots,K\}/\bi$ dimensions of the histogram $\bS$. Hence, $\bS^\bi$ is truly a marginal histogram of $\bS$ in the usual sense of the term.

Similarly to \eqref{histlik}, the resulting symbolic likelihood function for the marginal histogram $\bS^\bi$
is then given by
\begin{equation}
L(\bS^{\bi};\theta) \propto \frac{N!}{s_{\bold 1^{\bi}}^{\bi}!\cdots s_{\bB^{\bi}}^{\bi}!}\prod_{\bb^{\bi}=\bold 1^{\bi}}^{\bB^{\bi}}P_{\bb^{\bi}}(\theta)^{s_{\bb^{\bi}}^{\bi}},
\label{eq:genS}
\end{equation}
where 
$\bs^{\bi}=(s_{\bold 1^{\bi}}^{\bi},\ldots,s_{\bB^{\bi}}^{\bi})$ denotes the observed value of $\bS^\bi$
and
$P_{\bb^{\bi}}(\theta) = \int_{\Upsilon_{\bb^\bi}}g_{\bX^\bi}^\bi(\bz^\bi;\theta)d\bz^\bi$ is the probability of observing a datapoint within the $I$-dimensional marginal bin $\Upsilon_{\bb^\bi}$ under the marginal model $$g_{\bX^\bi}^\bi(\bx^\bi;\theta)=\int g_\bX(\bz;\theta)d\bz^{-\bi},$$ where $\bz^{-\bi}$ is the vector of elements of $\bz$ that are not in $\bz^\bi$.
In the case where $I=\{1,\ldots,K\}$ then \eqref{eq:genS} is equal to \eqref{histlik}.

Following similar arguments to \cite{beranger+ls18}, the symbolic likelihood $L(\bS^\bi;\theta)$ approaches the equivalent classical data likelihood $L(\bX^\bi;\theta)=g_{\bX^\bi}^\bi(\bX^i;\theta)$ 
as the number of bins in the marginal histogram approaches infinity and the volume of each bin approaches zero.
In particular, suppose for simplicity that the length $|\Upsilon_\bb^k|=y^k_{b_k}-y^k_{b_{k}-1}$ of each univariate marginal bin $\Upsilon_\bb^k=(y^k_{b_{k}-1},y^k_{b_k}]$ is equal for each margin $k=1,\ldots,K$, with fixed endpoints $y^k_0$ and $y^k_{B^k}$. Then as $B^k\rightarrow\infty$ the number of equally spaced bins grows, but their length $|\Upsilon_\bb^k|\rightarrow 0$.
Then
\[
	\lim_{\substack{B^k\rightarrow\infty\\k=1,\ldots,K}}
	L(\bS^{\bi};\theta) = L(\bX^{\bi};\theta).
\]
Intuitively in this setting, as the number of bins gets large and their volume reduces, in the limit almost all bins will be empty, with each observed datapoint $x^\bi$
being contained in exactly one bin. For the symbolic likelihood \eqref{eq:genS}, this means that empty bins ($\bs^\bi_{\bb^\bi}=0$) will not contribute to the likelihood, and the $N$ non-empty bins ($\bs^\bi_{\bb^\bi}=1$) will contribute the term $g_{\bX^\bi}^\bi(\bx^\bi;\theta)=g_{X^\bi}(x^\bi;\theta)$, which is the equivalent term contributed to the  classical likelihood function $L(\bX^\bi;\theta)$.

 As a result, this means that taking more bins will allow $L(\bS^\bi;\theta)$, taken as an approximation to $L(\bX^\bi;\theta)$, to approximate the classical data likelihood arbitrarily well. The difference is that the symbolic likelihood contains $B^1\times\ldots\times B^K$ terms, which may be considerably less than the $N$ terms of the classical data likelihood $L(\bX^\bi;\theta) = \prod_{k=1}^Ng_{X^\bi}(x^\bi_k;\theta)$ for large datasets. In this setting, the tradeoff of improved computational efficiency for some, perhaps small, approximation error may be attractive.

In particular, we may construct the log-likelihood function of a bivariate random marginal histogram $\bS^{\bi_2}$ by specifying the indices $\bi_2=(i_1,i_2)$, marginal bin indices $\bb_2=(b_{i_1},b_{i_2})$ and number of bins $B^{i_1}\times B^{i_2}$, giving
\begin{equation}
	\ell(\bS^{\bi_2};\theta)  \propto \sum_{b_{i_1}=1}^{B^{i_1}}\sum_{b_{i_2}=1}^{B^{i_2}}{s_{(b_{i_1},b_{i_2})}^{\bi_2}}\log P_{(b_{i_1},b_{i_2})}(\theta).
	\label{eq:bivS}
\end{equation}
  Similarly, specifying $\bi = (i_1,i_2,i_3)$ leads to the log-likelihood function of a trivariate random marginal histogram $\bS^{\bi_3}$ with $B^{i_1}\times B^{i_2}\times B^{i_3}$ bins indexed by $\bb_3=(b_{i_1},b_{i_2},b_{i_3})$, given by
\begin{equation}
	\ell(\bS^{\bi_3};\theta)  \propto \sum_{b_{i_1}=1}^{B^{i_1}}\sum_{b_{i_2}=1}^{B^{i_2}}\sum_{b_{i_3}=1}^{B^{i_3}}{s_{(b_{i_1},b_{i_2},b_{i_3})}^{\bi_3}}\log P_{(b_{i_1},b_{i_2},b_{i_3})}(\theta).
	\label{eq:trivS}
\end{equation}
Clearly the number of terms  in the full symbolic likelihood \eqref{histlik}, $B^1\times\ldots\times B^K$, increases exponentially as the dimension of the histogram, $K$, increases. This is further compounded  since larger $B^k$, $k=1,\ldots,K$, will produce a closer likelihood approximation $L(\bS;\theta) \approx L(\bX;\theta)$, which may be desirable. Similarly, the complexity of efficiently computing the $K$-dimensional integral $$P_\bb(\theta)=\int_{\Upsilon_\bb} g_\bX(\bz;\theta)d\bz$$ also increases with $K$. Together this means that it may rapidly become practically infeasible to directly use the symbolic likelihood of \cite{beranger+ls18} in more than, say, $K=5$ or $6$ dimensions, which reduces the applicability of this approach. However, the computational overheads of the bivariate and trivariate marginal histogram log-likelihoods  \eqref{eq:bivS} and \eqref{eq:trivS} will be much lower. This motivates the use of composite likelihood techniques, constructed from marginal histograms $\bS^\bi$ of $\bS$, which we now describe within the symbolic likelihood setting.

\subsection{Composite likelihoods}
\label{ssec:1}

Composite likelihoods, part of the family of pseudo-likelihood functions, are one practical technique for constructing asymptotically consistent likelihood-based parameter estimates when the standard likelihood function is computationally intractable \citep{lindsay1988,varin+rf11}. Such intractability can occur in many common modelling scenarios \citep{varin+v05,sisson+fb18}.
In particular, in Section \ref{sec:examples} we examine max-stable process models for spatial extremes \citep{davison+pr12,padoan+rs10}, for which closed-form densities are available for models with $K=2$ or $3$ spatial locations, but not for the larger $K$ required in practical applications, typically measured in the hundreds. See Section \ref{sec:examples} for further details.
Composite likelihoods are defined as the weighted product of conditional or marginal events of a process, each of which may be described by e.g.~an ordinary likelihood function \citep{lindsay1988}. 
If we assume all weights are equal for simplicity, a composite likelihood function can be expressed as $L_{CL}(\bx;\theta) \propto \prod_{i=1}^mL_i(\bx;\theta)$, where $L_i(\bx;\theta)$ is the likelihood function of a conditional or marginal event of $\bx$ for a given parameter vector $\theta$. 

A special case of the composite likelihood function is the $j$-wise composite likelihood function, comprising all $j$-dimensional marginal events.
Using the same notation as in Section \ref{symbolicmodel}, and defining ${\mathcal I}_j=\{\bi: \bi\subseteq\{1,\ldots,K\}, |\bi|=j\}$ to be the set of all $j$-dimensional subsets of $\{1,\ldots,K\}$,
the $j$-wise composite likelihood function can be written as
\begin{equation}
\label{eqn:jwiseCmle}
L_{CL}^{(j)}(\bx;\theta) \propto \prod_{\bi\in{\cal I}_j} g^\bi_{\bX^\bi}(\bx^{\bi};\theta),
\end{equation}
where, as before, $g^\bi$ represents the $j$-dimensional (marginal) density associated with the $j$-wise event $\bi\in{\cal I}_j$. 
In analogy with \eqref{eq:bivS} and \eqref{eq:trivS},
when $j=2$ the pairwise composite log-likelihood function, $\ell_{CL}^{(2)}$, is given by
\begin{equation}
	\ell_{CL}^{(2)}(\bx;\theta) \propto \sum_{i_1=1}^{K-1}\sum_{i_2=i_1+1}^K\log g_{\bX^{i_1},\bX^{i_2}}(\bx^{i_1},\bx^{i_2};\theta),
	\label{eq:6}
\end{equation}
and similarly for $j=3$, the triple-wise composite log-likelihood, $\ell_{CL}^{(3)}$, is given by
\begin{align*}
	\ell_{CL}^{(3)}(\bx;\theta)& \propto \\
	 \sum_{i_1=1}^{K-2}&\sum_{i_2=i_1+1}^{K-1}\sum_{i_3=i_2+1}^K\log g_{\bX^{i_1},\bX^{i_2},\bX^{i_3}}(\bx^{i_1},\bx^{i_2},\bx^{i_3};\theta).
\end{align*}

Taking first order partial derivatives of $\ell_{CL}^{(j)}(\bx; \theta)$ with respect to $\theta$ yields the composite score function $\nabla \ell_{CL}^{(j)}(\theta;\bx)$, 
and taking second order partial derivatives gives the Hessian matrix $\nabla^2 \ell^{(j)}_{CL}(\theta;\bx)$.
\cite{lindsay1988} showed that the resulting maximum $j$-wise composite likelihood estimator, $\hat \theta_{CL}^{(j)}$, is asymptotically consistent and distributed as
$$
\sqrt N \left(\hat \theta_{CL}^{(j)}-\theta \right) \rightarrow N \left(0,G^{(j)}(\theta)^{-1} \right),
$$
where $G^{(j)}$ is the ($j$-wise) Godambe information matrix \citep{godambe1960} defined by $$G^{(j)}(\theta) = H^{(j)}(\theta)J^{(j)}(\theta)^{-1}H^{(j)}(\theta),$$ 
where $H^{(j)}(\theta) = -\mathbb E_g(\nabla^2\ell_{CL}^{(j)}(\theta;\bx))$ and $J^{(j)}(\theta) = \mathbb V_g(\nabla\ell_{CL}^{(j)}(\theta;\bx))$ are respectively the sensitivity and variability matrices.
For standard likelihoods we have $j=K$ and $\mathcal{I}=\{(1,\ldots,K)\}$, and so dropping the superscripts, 
$H(\theta)=J(\theta)$ and 
the Godambe information matrix reduces to $G(\theta) = H(\theta) = I(\theta)$, where $I(\theta)$ is the Fisher information matrix.
The above result shows that the composite MLE is asymptotically unbiased, however it is worth noting that $G(\theta)^{-1}$ often does not attain the Cramer-Rao lower bound and subsequently there is a decrease in efficiency when the composite MLE is used in the place of the standard MLE \citep{varin+rf11}.

{\color{black}
The number of terms in the $j-$wise composite likelihood function (\ref{eqn:jwiseCmle}) increases exponentially with an increasing number of dimensions $K$.
\cite{padoan+rs10} empirically demonstrated that for a max-stable process model with $K$ spatial dimensions (which we use here in 
Sections \ref{sec:examples} and \ref{sec:6}), the trace of the asymptotic covariance matrix of the estimate is minimised if most pairs are excluded from the likelihood, leaving only pairs within a low taper distance and gains in both efficiency and computation. \cite{sang2013} consider the use of tapering to reduce this computational burden, giving each term in the likelihood (\ref{eqn:jwiseCmle}) a weight 1 for all sets of pairs or triples located closer than a specified taper distance, and 0 otherwise. The optimal value for the taper distance is obtained via the minimisation of two different characteristics of the covariance matrix; the trace and the determinant. The computational cost of obtaining the optimal taper distance can be minimised by using a subsample of the data.

\cite{bevil2012} propose two approaches to estimate covariance functions for space and space-time data. The first bases the weights on the distance between the pairs/triples, with a maximal value of 1, and 0 for all locations further apart than a given taper distance. The second method chooses weights by minimising the asymptotic covariance matrix of the model parameters. Both methods are empirically shown to significantly outperform the equal-weights approach in terms of efficiency and computational burden. \cite{li2018} derive optimal weights for each tuple in the $j-$wise likelihood of a Gaussian process by choosing them to obtain the optimal estimating equations for the model. Tapering and the assumption of a block diagonal form for the sets of locations are then used to reduce computational costs.

Such weighted composite likelihood methods reduce the computational burden associated with increasing dimension $K$, but can still be computationally infeasible for datasets with large numbers of observations $N$. In the following Section, we introduce a histogram-based approach to reduce the computational burden associated with the composite likelihood analysis of large datasets. While we do not pursue it here,  the tapering and related methods described above can be implemented in this setting to further improve computational efficiency.

}

\section{Composite likelihood functions for histogram-valued data}
\label{sec:compsym}


In this Section we introduce a composite likelihood function for random histograms that is constructed using sets of marginal histograms. We will first present the main result, before examining the consistency and variability of the symbolic composite MLE in turn, as the form of each of these has interesting implications for statistical inference using random histograms. 

\subsection{Composite likelihood function}
Suppose that we observe $T$ independent replicates, $\bX_1,\ldots,\bX_T$,  of the random variable $\bX=(X_1,\ldots,X_N) \in \mathbb R^{K\times N}${\color{black}{ over}} some index variable $t=1,\ldots, T$, and denote the realised values as $\bx_t$. For each $\bX_t$, $t=1,\ldots,T$, we may construct a $K$-dimensional random histogram $\bS_t$ over the set of bins $\{\boldsymbol{1},\ldots,\bB\}$. A $j$-dimensional marginal histogram of $\bS_t$ may then be constructed as $\bS_t^\bi$, where $\bi\in{\mathcal I}_j$. For a given model $g_\bX(\bx;\theta)=\prod_{i=1}^Ng_X(x_i;\theta)$ for the micro-data $\bX_t$, the likelihood of the marginal histogram $\bS_t^\bi$ is then given by $L(\bS_t^\bi;\theta)$ in \eqref{eq:genS}.
We can now define the $j$-wise symbolic composite likelihood for all $j$-dimensional marginal histograms $\bS_t^\bi$ of $\bS_t$, $i\in{\mathcal I}_j$, $t=1,\dots, T$ as follows.

\begin{mydef}
\label{propsl}
Writing $\bS_{1:T}=(\bS_1,\ldots,\bS_T)$ as the collection of $K$-dimensional histograms, the $j$-wise symbolic composite likelihood for $\bS_{1:T}$ is given by
\begin{equation}
L_{SCL}^{(j)}(\bS_{1:T};\theta) = \prod_{t=1}^T\prod_{\bi\in{\mathcal I}_j}L(\bS_{t}^{\bi};\theta),
\label{eq:4old}
\end{equation}
where $L(\bS_{t}^{\bi};\theta)$ is defined in \eqref{eq:genS}. 
Defining the maximum $j$-wise symbolic composite likelihood estimator as $\hat{\theta}^{(j)}_{SCL}=\arg\max_\theta L_{SCL}^{(j)}(\bS_{1:T};\theta)$, following standard composite likelihood construction arguments \citep{lindsay1988} we have
\[
	{\color{black}{\sqrt{T}}}\left(\hat{\theta}^{(j)}_{SCL}-\theta\right)\rightarrow {\mathcal N}\left(0,G^{(j)}(\theta)^{-1}\right),
\]
as ${{\color{black}T}\rightarrow\infty}$ where $G^{(j)}(\theta)=H^{(j)}(\theta)J^{(j)}(\theta)^{-1}H^{(j)}(\theta)$, and where estimates of the sensitivity and variability matrices are given by
\label{propg}
\begin{align}
\label{eqn:Hhat} \hat H(\hat{\theta}_{SCL}^{(j)}) &= -\sum_{t=1}^T\sum_{\bi\in{\mathcal I}_j}\nabla^2\ell(\bS_{t}^{\bi};\theta)\\
& = -\sum_{t=1}^T\sum_{\bi\in{\mathcal I}_j}
\sum_{\bb^{\bi}=\boldsymbol{1}^\bi}^{\bB^{\bi}}
s_{t,\bb^{\bi}}^{\bi}\nabla^2\log P_{t,\bb^{\bi}}(\hat \theta_{SCL}^{(j)})\\
\hat J(\hat{\theta}_{SCL}^{(j)}) &= \sum_{t=1}^T\left (\sum_{\bi\in{\mathcal I}_j}\nabla \ell(\bS_{t}^{\bi};\theta)\right )\left (\sum_{\bi\in{\mathcal I}_j}\nabla \ell(\bS_{t}^{\bi};\theta)\right )^\top\nonumber\\
\label{eqn:Jhat} & = \sum_{t=1}^T\left (\sum_{\bi\in{\mathcal I}_j}\sum_{\bb^{\bi}=\boldsymbol{1}^\bi}^{\bB^{\bi}}s_{t,\bb^{\bi}}^{\bi}\nabla\log P_{t,\bb^{\bi}}(\hat \theta_{SCL}^{(j)})\right )\times \\
&\left (\sum_{\bi\in{\mathcal I}_j}\sum_{\bb^{\bi}=\boldsymbol{1}^{\bi}}^{\bB^{\bi}}s_{t,\bb^{\bi}}^{\bi}\nabla\log P_{t,\bb^{\bi}}(\hat \theta_{SCL}^{(j)})\right )^\top,
\end{align}
where $t$ subscripts indicate dependence on $\bS_t$.
\end{mydef}

For example, the pairwise ($j=2$) symbolic composite log-likelihood function is given by
\begin{equation}
\ell_{SCL}^{(2)}(\bS_{1:T};\theta) = \sum_{t=1}^T\sum_{i_1=1}^{K-1}\sum_{i_2=i_1+1}^K\ell(\bS_{t}^{(i_1,i_2)};\theta)
\label{eq:4}
\end{equation}
where $\ell(\bS_{t}^{(i_1,i_2)};\theta)$ is given by \eqref{eq:bivS}, 
and the triple-wise ($j=3$) symbolic composite log-likelihood function is given by
\begin{equation}
\ell_{SCL}^{(3)}(\bS_{1:T};\theta) = \sum_{t=1}^T\sum_{i_1=1}^{K-2}\sum_{i_2=i_1+1}^{K-1}\sum_{i_3=i_2+1}^{K}\ell(\bS_{t}^{(i_1,i_2,i_3)};\theta)
\label{eq:5}
\end{equation}
where $\ell(\bS_{t}^{(i_1,i_2,i_3)};\theta)$ is given by \eqref{eq:trivS}.

\subsection{Symbolic composite maximum likelihood estimator consistency} 
\label{ssec:a2}

It is straightforward to show that the $j$-wise symbolic composite likelihood estimator $\hat{\theta}_{SCL}^{(j)}$ that maximises \eqref{eq:4old} is consistent with the equivalent composite likelihood estimator $\hat{\theta}_{CL}^{(j)}$ that maximises $$L_{CL}^{(j)}(\bX_{1:T};\theta)=\prod_{t=1}^T L_{CL}^{(j)}(\bX_t;\theta)$$ where $L_{CL}^{(j)}(\bX_t;\theta)$ is given by \eqref{eqn:jwiseCmle} as the number of bins in each marginal histogram approaches infinity and the volume of each bin approaches zero.

We show this by extending the univariate proof described by \cite{zhang2017} to (w.l.o.g) the bivariate ($j=2$) setting, from which the extension to the $K$-dimensional case is immediate.

Consider the pairwise composite log likelihood given in \eqref{eq:4}. In this case, for $\bi=(i_1,i_2)\in{\mathcal I}_2$, and for any $t=1,\ldots,T$ (although dropping the subscript $t$ for clarity), the probability that a bivariate micro-data observation $X^\bi\in\mathbb{R}^2$ falls in marginal bin $\bb^\bi=(b_{i_1},b_{i_2})$ over the region $ ( y^{i_1}_{b_{i_1}-1}, y^{i_1}_{b_{i_1}} ]\times ( y^{i_2}_{b_{i_2}-1}, y^{i_2}_{b_{i_2}} ]$ is
\begin{align*}
	P_{\bb^{\bi}}(\theta) &= G^\bi_{X^\bi}(y^{i_1}_{b_{i_1}},y^{i_2}_{b_{i_2}};\theta)-G^\bi_{X^\bi}(y^{i_1}_{b_{i_1}-1},y^{i_2}_{b_{i_2}};\theta)\\
	&-G^\bi_{X^\bi}(y^{i_1}_{b_{i_1}},y^{i_2}_{b_{i_2}-1};\theta)+G^\bi_{X^\bi}(y^{i_1}_{b_{i_1}-1},y^{i_2}_{b_{i_2}-1};\theta),
\end{align*}
where $G_X(x;\theta)$ is the distribution function of $g_X(x;\theta)$.

Fixing the $i_2$ margin, by the mean value theorem there exists a $\tilde x_{b_{i_1}} \in (y^{i_1}_{b_{i_1}-1},y^{i_1}_{b_{i_1}}]$ such that
\begin{align*}
	P_{\bb^{\bi}}(\theta) 
	=&(y^{i_1}_{b_{i_1}}-y^{i_1}_{b_{i_1}-1})\frac{d}{dx_1}G^\bi_{X^\bi}(\tilde x_{b_{i_1}}, y^{i_2}_{b_{i_2}};\theta)\\
	&-(y^{i_1}_{b_{i_1}}-	y^{i_1}_{b_{i_1}-1})\frac{d}{dx_1}G^\bi_{X^\bi}(\tilde x_{b_{i_1}}, y^{i_2}_{b_{i_2}-1};\theta),
\end{align*}
where $\frac{d}{dx_k}G_X$ denotes differentiation with respect to the $k$-th component of $G_X$.
Similarly fixing the $i_1$ margin, again by the mean value theorem there exists a $\tilde x_{b_{i_2}} \in (y^{i_2}_{b_{i_2}-1},y^{i_2}_{b_{i_2}}]$ such that
\begin{align*}
P_{\bb^{\bi}}(\theta)&= (y^{i_1}_{b_{i_1}}-y^{i_1}_{b_{i_1}-1})\frac{d}{dx_1}\big[G_X(\tilde x_{b_{i_1}}, y^{i_2}_{b_{i_2}};\theta)\\
&-G_X(\tilde x_{i_{b_1}}, y^{i_2}_{b_{i_2}-1};\theta) \big]\\
&=(y^{i_1}_{b_{i_1}}-y^{i_1}_{b_{i_1}-1})(y^{i_2}_{b_{i_2}}-y^{i_2}_{b_{i_2}-1})\frac{d}{dx_1}\frac{d}{dx_2}G_X(\tilde x_{b_{i_1}}, \tilde x_{b_{i_2}};\theta)\\
& \propto \frac{d}{dx_1}\frac{d}{dx_2}G_X(\tilde x_{b_{i_1}}, \tilde x_{b_{i_2}};\theta) = g_X(\tilde x_{b_{i_1}}, \tilde x_{b_{i_2}};\theta).
\end{align*}
\textcolor{black}{Consequently, the pairwise symbolic composite log likelihood is given as}
\begin{align*}
	&\ell_{SCL}^{(2)}(\bS_{1:T};\theta)  \propto\\
	& \sum_{t=1}^T \sum_{i_1=1}^{K-1}\sum_{i_2=i_1+1}^K
	\sum_{b_{i_1}=1}^{B^{i_1}} \sum_{b_{i_2}=1}^{B^{i_2}}
	\bs^{\bi}_{(b_{i_1},b_{i_2})}
	\log g^\bi_{X^\bi}(\tilde{x}_{b_{i_1}}, \tilde {x}_{b_{i_2}};\theta).
\end{align*}
Now, letting the number of bins $B^{i_1}, B^{i_2}\rightarrow\infty$ such that each bin's volume $\rightarrow 0$ means that in the limit each bin will either contain zero ($s^\bi_{(b_{i_1},b_{i_2})}=0$) or, assuming continuous data, exactly one observation ($s^\bi_{(b_{i_1},b_{i_2})}=1$).
In the case where a bin contains exactly one observation, the $m$-th observed classical datapoint $(x_{m,i_1},x_{m,i_2})$, we have $(y^{i_1}_{b_{i_1}-1},y^{i_1}_{b_{i_1}}]  \times (y^{i_2}_{b_{i_2}-1},y^{i_2}_{b_{i_2}}]\rightarrow (x_{m,i_1},x_{m,i_2})$. Hence $(\tilde{x}_{b_{i_1}}, \tilde {x}_{b_{i_2}})\rightarrow (x_{m,i_1},x_{m,i_2})$ and so
\begin{align*}
	\ell_{SCL}^{(2)}&(\bS_{1:T};\theta)  \rightarrow\\
	& \sum_{t=1}^T \sum_{i_1=1}^{K-1}\sum_{i_2=i_1+1}^K
	\sum_{m=1}^N 
	\log g^\bi_{X^\bi}(x_{m,b_{i_1}}, x_{m,b_{i_2}};\theta),
\end{align*}
which is has a maximum at $\hat{\theta}^{(2)}_{CL}$.
This argument straightforwardly extends to the $j$-wise symbolic composite likelihood by iterated use of the mean value theorem.

This result means that the symbolic composite likelihood can be considered an asymptotically (in the number of bins) consistent approximation of the standard composite likelihood{, \color{black}{
which itself provides an asymptotically (in the number of datapoints) consistent estimator of the true parameter.
That is, as the number of bins increases (and the bin volume decreases), then the symbolic composite MLE approaches the standard composite MLE $\hat{\theta}^{(j)}_{SCL}\rightarrow\hat{\theta}^{(j)}_{CL}$. Further, if in addition the amount of data $N$ increases, then as $\hat{\theta}^{(j)}_{CL}\rightarrow\theta_0$ (where $\theta_0$ is the true parameter value) then $\hat{\theta}^{(j)}_{SCL}\rightarrow\theta_0$.
The symbolic composite approximation}} can be arbitrarily close {\color{black}{to the classical composite equivalent}} (although at the cost of increasing computational overheads) as the number of bins increases.


There are a number of specifications under which the $T$ random histograms may be constructed from the underlying micro-data (and the details of these are encoded in the parameter $\phi$ in \eqref{eq:1}). These specifications control the location and sizes of the bins in each random histogram, and the number of random histograms, $T$, itself. 
While we do not discuss the merits of particular constructions here, we note that the above asymptotic consistency result for the symbolic composite log likelihood holds regardless of the method of bin construction in each histogram (as long as the volume of each bin approaches zero as the number of bins approaches infinity), and regardless of the number of random histograms, $T$ (as long as the underlying micro-data $X_1,\ldots,X_N$ are stationary). Consistency also holds for different numbers of micro-data encoded in each random histogram $\bS_t$ as long as there is sufficient data in enough unique bins that $\ell(\bS_t^\bi;\theta)$ is well defined and satisfies the usual regularity conditions.

In particular, if each random histogram has exactly the same bins, so that $y^k_{t,b_k}=y^k_{b_k}$ for all $t=1,\ldots,T$, then the choice of $T$ has no effect on the symbolic composite maximum likelihood estimator. That is, $\hat{\theta}_{SCL}$ takes the same value independently of the number of random histograms $T$. This is easily seen as
\begin{equation}
\label{eqn:sum2s}
	\sum_{t=1}^Ts_{t,\bb^{\bi}}^{\bi} = s_{\bb^{\bi}}^{\bi},\mbox{ } \forall  \bb, \bi,
\end{equation}
where $s_{\bb^{\bi}}^{\bi}$ is the count of all micro-data falling in (marginal) bin $\bb^\bi$ when all data are allocated to a single ($T=1$) histogram. As a result, we then have
$$\sum_{t=1}^T\sum_{\bi\in{\mathcal I}_j}\sum_{\bb^{\bi}=\boldsymbol{1}^\bi}^{\bB^{\bi}}s_{t,\bb^{\bi}}^{\bi}\log P_{t,\bb^{\bi}}( \theta) = \sum_{\bi\in{\mathcal I}_j}\sum_{\bb^{\bi}=\boldsymbol{1}^\bi}^{\bB^{\bi}}s_{\bb^{\bi}}^{\bi}\log P_{\bb^{\bi}}( \theta),$$
and so the resulting symbolic composite maximum likelihood estimators are equivalent.
As a result, if primary interest of an analysis is of fast computation of $\hat{\theta}_{SCL}$, then the optimal choice is by constructing $T=1$ random histograms, as this will allow for the fastest optimisation of $\ell^{(j)}_{SCL}(\bS_{1:T};\theta)$. (Note that if all bins are equal, then this single histogram can be created by simply summing the counts in each bin, following \eqref{eqn:sum2s}.) However, $T=1$ will not be the optimal choice if interest is also in computing $\mbox{Var}(\hat{\theta}^{(j)}_{SCL})$ -- see the following Section.

\subsection{Variance consistency}
\label{ssec:999}

We now show the conditions under which the symbolic Godambe information matrix $G(\hat \theta_{SCL}^{(j)})$ converges  to the standard Godambe matrix $G(\hat \theta_{CL}^{(j)})$.
In particular, we will show that  as the number of equally spaced histogram bins becomes large (so that $B^k\rightarrow\infty$ for $k=1,\ldots,K$) while the volume of each bin approaches zero
 ($|\Upsilon_\bb|\rightarrow 0$, $\forall \bb$),
and as the number of histograms $T\rightarrow N$ so that each histogram contains exactly one micro-data observation, then
\[
	\lim_{T \rightarrow N}
	\lim_{\substack{B^k\rightarrow\infty\\k=1,\ldots,K}}
	\mbox{Var} (\hat \theta_{SCL}^{(j)}) = \mbox{Var}(\hat \theta_{CL}^{(j)}).
\]
 
 Following the same arguments as in Section \ref{ssec:a2} it is straightforward to show that
 \[
 	\lim_{\substack{B^k\rightarrow\infty\\k=1,\ldots,K}}
	\hat{H}(\hat{\theta}_{SCL}^{(j)}) = \hat{H}(\hat{\theta}_{CL}^{(j)}),
\]
so that the symbolic Hessian matrix converges to the standard composite likelihood Hessian matrix, regardless of the number of histograms, $T$, due to the additive form of \eqref{eqn:Hhat}. 
Numerical estimates of $\hat{H}(\hat{\theta}_{SCL}^{(j)})$ can be obtained through numerical methods during maximum likelihood estimation (e.g.~using the {\tt optim} function in {\tt R}).

 The natural estimator for the variability matrix is the empirical variance estimator \eqref{eqn:Jhat}.
With increasing $T$, the sum of the counts in each  histogram $\bS_{t}$ decreases in magnitude until there is exactly $1$ non-empty bin with count $1$ in each of $T=N$  marginal histograms. At this point 
\[
	\sum_{\bb^{\bi}=\boldsymbol{1}^\bi}^{\bB^{\bi}}s_{t,\bb^{\bi}}^{\bi}=1, \qquad \forall \bi\in{\mathcal I}_j,\: t=1,\ldots,N.
\]
As a result, the limit of the symbolic composite log-likelihood function, as $T\rightarrow N$, is
\begin{align*}
	\lim_{T\rightarrow N}&\ell_{SCL}^{(j)}(\bS_{1:T}; \theta)  \\
	&\propto\lim_{T\rightarrow N} \sum_{t=1}^T \sum_{\bi\in{\mathcal I}_j} \sum_{\bb^\bi=\boldsymbol{1}^\bi}^{\bB^\bi}  \bs^\bi_{t,\bb^\bi}\log P_{t,\bb^\bi}(\theta)\\
	&=  \sum_{t=1}^N \sum_{\bi\in{\mathcal I}_j} \log P_{t,\bb^{(t)\bi}}(\theta),
\end{align*}
where $\bb^{(t)}$ denotes the bin which contains the single micro-data observation $x_t$ in histogram $\bS_t$.
Because 
\[
	\lim_{\substack{B^k\rightarrow\infty\\k=1,\ldots,K}}\log P_{t,\bb^{(t)\bi}}(\theta) = \log g_{X^\bi}^\bi(x_t^{\bi};\theta)
\]
reduces to the standard composite likelihood marginal event component as the histogram bins reduce in size, then $\lim_{\substack{B^k\rightarrow\infty\\k=1,\ldots,K}}\hat{\theta}^{(j)}_{SCL}=\hat{\theta}^{(j)}_{CL}$.
It then follows that from \eqref{eqn:Jhat}
\begin{align*}
	&\lim_{T\rightarrow N}\lim_{\substack{B^k\rightarrow\infty\\k=1,\ldots,K}} \hat{J}(\hat{\theta}^{(j)}_{SCL})\\
	=&\lim_{\substack{B^k\rightarrow\infty\\k=1,\ldots,K}}\sum_{t=1}^N\left(\sum_{\bi\in{\mathcal I}_j} \nabla P_{t,\bb^{(t)\bi}}(\hat{\theta}^{(j)}_{SCL})\right)\\
	\times &\left(\sum_{\bi\in{\mathcal I}_j} \nabla P_{t,\bb^{(t)\bi}}(\hat{\theta}^{(j)}_{SCL})\right)^\top\\
	=&\lim_{\substack{B^k\rightarrow\infty\\k=1,\ldots,K}}\sum_{t=1}^N\left(\sum_{\bi\in{\mathcal I}_j} \nabla g_{X^\bi}^\bi(x_t^{\bi};\hat{\theta}^{(j)}_{CL})\right)\\
	\times &\left(\sum_{\bi\in{\mathcal I}_j} \nabla g_{X^\bi}^\bi(x_t^{\bi};\hat{\theta}^{(j)}_{CL})\right)^\top\\
	 = & 
	\hat{J}(\hat{\theta}^{(j)}_{CL}).
\end{align*}
Convergence of the symbolic Godambe information matrix $G(\hat \theta_{SCL}^{(j)})$ to the standard Godambe matrix $G(\hat \theta_{CL}^{(j)})$ then follows under these limit conditions.

 While the above result confirms that the limiting behaviour of $\hat{\theta}^{(j)}_{SCL}$ is the same as $\hat{\theta}^{(j)}_{CL}$, in particular as $T\rightarrow N$, in practice we may prefer to have less than $N$ random histograms for a given analysis, particularly if $N$ is very large. In this setting, for a fixed $T<N$ we then have
  \begin{align*}
  \label{classicgodambe}
	\lim_{\substack{B^k\rightarrow\infty\\k=1,\ldots,K}}& \hat{J}(\hat{\theta}^{(j)}_{SCL})\\
	= &
	\sum_{t=1}^T\left(\sum_{\bi\in{\mathcal I}_j} \nabla g_{\bX^\bi}^\bi(\bx_t^{\bi};\hat{\theta}^{(j)}_{SCL})\right)\left(\sum_{\bi\in{\mathcal I}_j} \nabla g_{\bX^\bi}^\bi(\bx_t^{\bi};\hat{\theta}^{(j)}_{SCL})\right)^\top
\end{align*}
using similar arguments to the above.

Compared to the standard composite likelihood sensitivity matrix $\hat{J}(\hat{\theta}^{(j)}_{CL})$, \eqref{classicgodambe} 
can be interpreted as the sensitivity matrix for a classical (micro-data) dataset where some temporal information is lost. That is, we know which time block (histogram) $t=1,...,T$ each observation came from, but not specifically when each observation occurred within that block. As a result the variability of $\hat{\theta}^{(j)}_{SCL}$ will always be larger for a smaller number of time blocks.
As $T$ increases, more temporal information is retained as 
each time block then decreases in size. This leads to more precise knowledge about when each data point may have been observed, and accordingly leading to a reduction in the variance of $\hat{\theta}^{(j)}_{SCL}$.
The standard composite likelihood case is recovered for $T=N$ when the time of each datapoint is known exactly. 

Equation \eqref{classicgodambe} thereby characterises the loss in precision for the standard composite MLE as temporal information is lost. It also characterises the limiting performance (in the sense of $B^k\rightarrow\infty$, $\forall k$) of the symbolic composite MLE. (This relationship is explored explicitly in Section \ref{ssec:65}.) However the advantage of working with $\hat{\theta}^{(j)}_{SCL}$ is that the likelihood function is typically more computationally efficient to evaluate for large $N$. As such, estimating $\mbox{Var}(\hat{\theta}^{(j)}_{SCL})$ represents a trade-off between greater precision (larger $T$) and greater computational and data storage efficiency (smaller $T$).

In practice, the analyst would choose $T$ as small as possible such that the inferential goals (perhaps depending on confidence intervals of model parameters) are still viable, in order to maximise overall analysis efficiency. Recall that, as discussed in Section \ref{ssec:a2}, if all histogram bins are equal, computation of the symbolic composite MLE itself can be achieved at low cost by combining all histograms into a single histogram ($T=1$). So the main impact of the number of histograms is on the variability of the symbolic composite MLE. 

{\color{black}{If the underlying micro-data are available, then $T=N$ histograms (again, with the same bins as for computing the symbolic composite MLE) can be used to determine the lowest possible variance of the  MLE. This is viable as the computational cost of evaluating the variance, given the MLE, is only a small fraction of the total computation required to optimise the likelihood (with $T=1$).}}

{\color{black}{{In terms of identifying the number of bins within each histogram, here we follow the strategy of increasing the value of $B$ sequentially in order to determine the point at which comparable estimates and standard errors to the classical composite likelihood function are obtained. For the value $B$ at which the change in results compared to the previous value of $B$ is negligible, the practitioner can be confident that further increasing the number of bins will not significantly improve the analysis, although it will increase the computational burden. While this approach is slightly inefficient as it requires repeated likelihood optimisation to compute multiple symbolic composite likelihood estimators $\hat{\theta}^{(j)}_{SCL}$ for varying values of $B$, the simulations in Section \ref{sec:examples} (e.g.~Table \ref{tab:tabletimes}) will demonstrate that this is still more computationally efficient than the existing classical analysis for large datasets, due to the large computational gains associated with employing a histogram-based approach. This simple approach is used in both the simulation studies in Section \ref{sec:examples}, and the real data analysis in Section \ref{sec:6}. }}}

\section{Simulation studies}  
\label{sec:examples}

We now examine the performance of the symbolic composite maximum likelihood estimator within the context of our motivating application -- modelling spatial extremes using max-stable processes. We first briefly introduce these, before comparing $\hat{\theta}^{(j)}_{SCL}$ to standard composite likelihoods in accuracy, precision and efficiency under a range of modelling scenarios.

\subsection{Max-stable process models}
\label{ssec:ext}

\cite{Jenkinson1955} first proposed a limiting distribution for modelling datasets comprising of block maxima. Suppose $X_1,\ldots,X_n\in D$, in some continuous space $D$, are i.i.d.~univariate random variables with distribution function $F$, and $$M_n=\max\{X_1,\ldots,X_n\}.$$ If there exist constants $a_n>0 $, $b_n\in \mathbb R$ such that
$$\lim_{n\rightarrow \infty}P\left( \frac{M_n-b_n}{a_n}\leq x\right) =\lim_{n \rightarrow \infty}F^n(a_nx+b_n)=G(x),$$
is non-degenerate, for all $x \in D$, then $G$ is a member of the generalised extreme value (GEV) family whose distribution function is given by
$G(x; \mu, \sigma, \xi) = \exp\{-v(x;\mu,\sigma,\xi)\},$
where $\mu \in \mathbb R$, $\sigma >0$, $\xi \in \mathbb R$, $v(y;\mu,\sigma,\xi) = \Big(1+\xi\frac{y-\mu}{\sigma}\Big)^{-\frac{1}{\xi}}_+$ when  $\xi \neq 0$ and $e^{-\frac{y-\mu}{\sigma}}$ otherwise, and $a_+ = \mbox{min}\{0,a\}.$ 


Max-stable processes \citep{dehaan1984,resnick1987,dehaan2006} are a popular tool to model spatial extremes.
Let $X_1, X_2, \ldots$ be a sequence of i.i.d.~copies of a stochastic process
$\{X(t): t\in \mathcal{T}\}$ over some space $\mathcal{T}$.
If continuous functions $a_n(t)>0$, $b_n(t)\in \mathbb R$ exist such that 
$$
\lim_{n\rightarrow \infty}  \frac{\max_{i=1,\ldots,n }X_i(t) -b_n(t)}{a_n(t)} =Y(t)
$$
is non-degenerate, then $Y(t)$ is a max-stable process. Spectral representations \citep{dehaan1984,schlather2002} allow to define max-stable models for $Y(t)$ such as the flexible extremal skew-$t$ \citep{beranger2017} and its particular cases. Here we select the Gaussian max-stable process \citep{Smith1990}, one of the simplest parametric models.
\cite{Genton2011} derived the joint  distribution function of this model for $K \geq 2$ spatial locations with coordinates $t_k \in \mathcal{T}=\mathbb R^d$, $k=1,\ldots,K$, where $K\leq d+1$.
Let $\tilde{T} = (t_1,\ldots,t_K)\in \mathbb R^{d\times K}$ be the matrix of coordinates for the locations, and $\tilde{T}_{-k}$ be the matrix $\tilde{T}$ without the $k^{th}$ column, $k=1,\ldots,K$.
Also let $\bold v = (v_1,\ldots,v_K)^\top \in \real^K_+$  and $c^{(j)}(\bold v) = \left(c_1^{(j)}(\bold v),\ldots,c_{j-1}^{(j)}(\bold v),c_{j+1}^{(j)}(\bold v),\ldots,c_K^{(j)}(\bold v) \right)^\top \in \real^{K-1}$, 
where, for $k=1,\ldots,K$,  $v_k = v(y_k;\mu,\sigma,\xi)^{-1}$ and $c_k^{(j)}(\bold  v) = (t_j-t_k)^\top\Sigma^{-1}(t_j-t_k)/2 -\log \left ( \frac{v_j}{v_k}\right)$.
Then, writing
$\Sigma^{(j)} = (t_j1_{K-1}^\top-\tilde{T}_{-j})^\top\Sigma^{-1}(t_j1_{K-1}^\top-\tilde{T}_{-j})$, where $1_d=(1,\ldots,1)^\top \in \mathbb R^d$,
the distribution function of the Gaussian max-stable process model can be written as 
\begin{align}
&P(Y_{1}(t)\le y_1, \ldots,Y_{K}(t) \le y_K)\\
= & \exp \left\{-\sum_{j=1}^K\frac{1}{v_j}\Phi_{K-1} \left(c^{(j)}(\bold v);\Sigma^{(j)} \right) \right\},
\label{blibblob}
\end{align}
where $\Phi_d(\,\cdot\,;\Sigma)$ is the $d-$dimensional zero-mean Gaussian distribution function with covariance matrix $\Sigma$. Each univariate margin of this process is a GEV distribution.
The parameters for this model are the spatial covariance matrix $\Sigma=[\sigma_{ij}]$ and the marginal GEV parameters $\mu,\sigma,\xi$.

For typical spatial problems the number of spatial locations $K$ is in the order of hundreds.
We use $K\sim100$ in some of the below simulations and the future-simulation climate data analysis in Section \ref{sec:6}. {\color{black} The complexity of the density function associated with \eqref{blibblob} and its computational burden  explodes with an increasing number of locations $K$, and consequently the complete likelihood isn't feasible in the analysis of such datasets. }
For this reason, composite likelihood techniques are attractive in practice.

In the following we compare the performance of both symbolic composite and standard composite likelihood  MLEs ($\hat{\theta}^{(j)}_{SCL}$ and $\hat{\theta}^{(j)}_{CL}$ respectively) in scenarios following those in  \cite{padoan+rs10} and \cite{Genton2011}, where $\theta=(\sigma_{11},\sigma_{12},\sigma_{22},\mu,\sigma,\xi).$
For each experiment, $K$ locations are generated uniformly over the space ${\mathcal T}=[0,40]\times[0,40]$  ($d=2$). For each location, $N$ realisations are generated from the Gaussian max-stable model using the {\tt R} package $\tt SpatialExtremes$ \citep{spatialextremes} with standard Gumbel margins (i.e.~$(\mu, \sigma, \xi) = (0,1,0)$).
 
\begin{figure}[h]
\centering
\includegraphics[width=\textwidth]{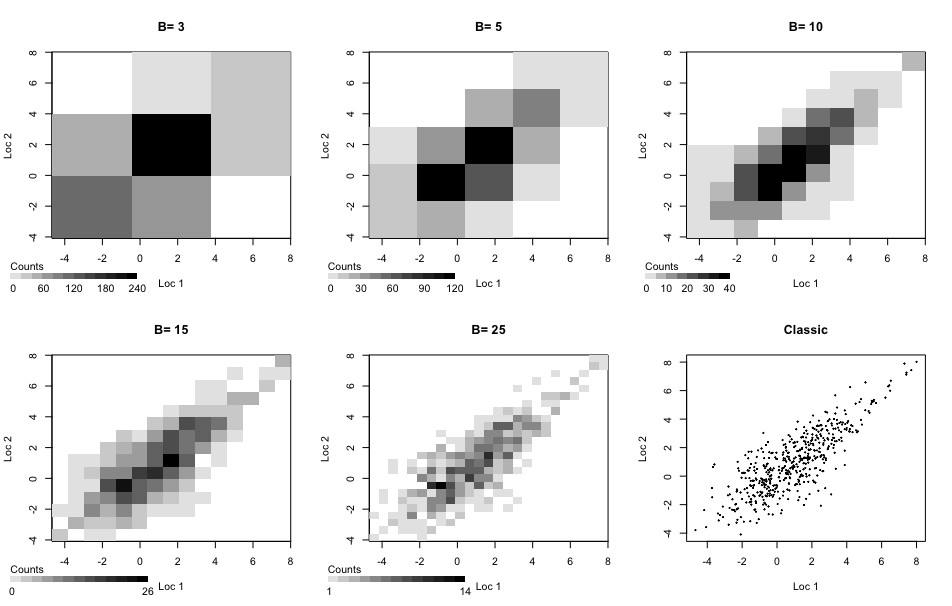}
 \caption{$B\times B$ bivariate histograms for different values of $B$ for the same classical dataset  (bottom right panel) of size $N=1\,000$, generated at two spatial locations under the Gaussian max-stable model with 
  $\Sigma=\Sigma_3$ (Table \ref{tab:true}). }
\label{fig:1}
\end{figure}

\begin{table}[h]
\small
\begin{center}
\begin{tabular}{ |c|c|c|c| } 
\hline
Model & $\sigma_{11}$&$\sigma_{12}$&$\sigma_{22}$\\
\hline
$\Sigma_1$ & 300 &0&300 \\ 
$\Sigma_2$ & 300 &150&300 \\ 
$\Sigma_3$ & 300 &150&200 \\ 
$\Sigma_4$ & 3\,000 &1\,500&3\,000 \\ 
$\Sigma_5$ & 30 &15&30 \\ \hline

\hline
\end{tabular}
\end{center}
 \caption{Spatial dependence parameter specifications
 for the Gaussian max-stable model, following \cite{padoan+rs10}.}
\label{tab:true}
\end{table}

\subsection{Comparisons with composite likelihoods}

\subsubsection{Varying the number of bins, $B$}

\label{ssec:61}
We generate $N=1\,000$ realisations for $K=15$ locations and $5$ different configurations of the covariance matrix $\Sigma$, with true values given in Table \ref{tab:true}, which represent a range of dependence scenarios. For each dataset a single histogram $\bS$ ($T=1$) is `constructed', although in practice we only construct all histograms $\bS^\bi, \bi\in{\mathcal I}_2$ for each pair of spatial locations.
The number of bins is constant in each dimension $B^k=B$, $k=1,\ldots,K$, and we specify $B=2, 3, 5, 10, 15$ and $25$. Figure \ref{fig:1} shows the resulting bivariate histograms for two locations with $\Sigma=\Sigma_3$.

Tables \ref{tab:table1} and \ref{tabex:table1} report the resulting mean symbolic composite and composite MLEs, $\hat{\theta}^{(2)}_{SCL}$ and $\hat{\theta}^{(2)}_{CL}$, with standard errors in parentheses, based on 1\,000 replicate analyses, for different values of $B$. 
While for low $B$ there is high variability in the estimates, as $B$ increases the mean MLEs and standard errors approach the same quantities obtained under the classical data analysis,  even in cases of very strong ($\Sigma_4$) or very weak ($\Sigma_5$) dependence.

In this case, comparable estimates to the composite MLEs are available for $B=25$, however practically viable estimates (with larger variances) can be obtained for much smaller values ($B\approx 10$).

\begin{table}[h]
\color{black}
\small
\centering
\begin{adjustbox}{width=0.6\textwidth}
\begin{tabular}{ |c|c|ccc| } 
\hline
Model & $B$ & $\sigma_{11}$&$\sigma_{12}$&$\sigma_{22}$ \\
\hline
\multirow{7}{*}{$\Sigma_1$} 
& 2 & 335.5 (585.5) & \phantom{-}5.7 (232.2) & 317.2 (125.1) \\ 
& 3  & 301.0 (\phantom{-}34.5)& -0.1 (\phantom{-}16.9) & 301.9 (\phantom{-}33.5) \\ 
& 5  & 299.1 (\phantom{-}23.1)&-0.9 (\phantom{-}13.2) & 299.9 (\phantom{-}24.1) \\ 
& 10& 299.8 (\phantom{-}20.2)&-0.5 (\phantom{-}11.1) & 300.0 (\phantom{-}20.9)\\ 
& 15& 299.8 (\phantom{-}18.9)&-0.3 (\phantom{-}10.4) & 300.0 (\phantom{-}19.5)\\ 
& 25& 299.7 (\phantom{-}18.0)&-0.3 (\phantom{-}10.0) & 300.2 (\phantom{-}18.9)\\ 
& \bf{Classic} &  {\bf 300.76 (17.1)} & {\bf -0.4 (9.7)} & {\bf 301.02 (18.1)}  \\ 
\hline
\multirow{7}{*}{$\Sigma_2$} 
& 2  & 316.59 (149.1)&165.1 (246.8) &332.9 (153.5) \\ 
& 3  & 299.6 (\phantom{-}35.0)&149.7 (\phantom{-}24.9)&300.8 (\phantom{-}33.7) \\
& 5  & 298.9 (\phantom{-}23.4)&149.2 (\phantom{-}16.7) & 299.9 (\phantom{-}23.4)\\ 
& 10& 299.3 (\phantom{-}20.2)&149.6 (\phantom{-}13.9) & 300.3 (\phantom{-}19.9)\\ 
& 15& 299.4 (\phantom{-}19.2)&149.7 (\phantom{-}13.2) & 300.5 (\phantom{-}19.0)\\ 
& 25& 299.7 (\phantom{-}18.3)&149.9 (\phantom{-}12.5) & 300.5 (\phantom{-}18.1)\\ 
& \bf{Classic} &  {\bf 300.7 (17.0)} & {\bf 150.4 (11.6)} & {\bf 301.53 (17.0)} \\ 
\hline
\multirow{7}{*}{$\Sigma_3$} 
& 2  & 321.6 (360.0) & 162.3 (210.6) & 210.8 (131.2) \\ 
& 3  & 296.1 (\phantom{-}30.6)&147.4 (\phantom{-}20.1)&197.9 (\phantom{-}19.9) \\ 
& 5  & 298.8 (\phantom{-}23.3)&149.4 (\phantom{-}15.3) & 199.6 (\phantom{-}15.4)\\ 
& 10& 299.0 (\phantom{-}19.3)&149.6 (\phantom{-}12.3) & 199.7 (\phantom{-}12.9)\\ 
& 15& 299.5 (\phantom{-}18.7)&149.8 (\phantom{-}11.6) & 199.8 (\phantom{-}12.1)\\ 
& 25& 299.7 (\phantom{-}17.8)&150.0 (\phantom{-}11.2) & 200.0 (\phantom{-}11.8)\\ 
& \bf{Classic} & {\bf 300.7 (16.4)} & {\bf 150.6 (10.2)} & {\bf 200.6 (10.9)}  \\ 
\hline
\multirow{7}{*}{$\Sigma_4$} 
& 2  & 3554 (2071)&1848 (1319) &3473 (1839) \\ 
& 3  & 2954 (\phantom{-}435)&1453 (\phantom{-}294)&2952 (\phantom{-}405) \\ 
& 5  & 3003 (\phantom{-}345)&1500 (\phantom{-}244) & 2996 (\phantom{-}337))\\ 
& 10& 3002 (\phantom{-}249)&1506 (\phantom{-}169) & 2997 (\phantom{-}239)\\ 
& 15& 2992 (\phantom{-}217)&1498 (\phantom{-}148) & 2988 (\phantom{-}211)\\ 
& 25& 2992 (\phantom{-}199)&1499 (\phantom{-}136) & 2991 (\phantom{-}200))\\ 
& \bf{Classic} &  {\bf 3002 (190)} & {\bf 1503 (124)} & {\bf 2999 (189)}  \\ 
\hline
\multirow{7}{*}{$\Sigma_5$} 
& 2  & 30.97 (3.57)&15.53 (2.81) &30.98 (3.86) \\ 
& 3  & 29.83 (2.04)&14.89 (1.58)&29.82 (2.18) \\ 
& 5  & 29.86 (1.54)&14.85 (1.17) & 29.82 (1.71)\\ 
& 10& 29.93 (1.27)&14.92 (0.95) & 29.91 (1.45)\\ 
& 15& 29.96 (1.20)&14.93 (0.91) & 29.91 (1.33)\\ 
& 25& 29.97 (1.13)&14.95 (0.86) & 29.94 (1.28)\\ 
& \bf{Classic} &  {\bf 30.10 (0.94)} &{\bf 15.06 (0.66)} & {\bf 30.06 (1.03)}  \\ 
\hline
\end{tabular}
\end{adjustbox}
 \caption{Mean (and standard errors) for the dependence parameters of the symbolic composite MLE $\hat{\theta}^{(2)}_{SCL}$ and composite MLE $\hat{\theta}^{(2)}_{CL}$ (Classic) from $1000$ replications of the Gaussian max-stable process model, for $B\times B$ histograms for varying values of $B$.
 Results based on $N=1\,000$ observations at $K=15$ spatial locations and $T=1$ random histogram.
}
\label{tab:table1}
\end{table}

\begin{table}[h]
\color{black}
\small
\centering
\begin{adjustbox}{width=0.5\textwidth}
\begin{tabular}{|c|c|ccc| } 
\hline
Model & $B$ &$\mu$&$\sigma$&$\xi$ \\
\hline
\multirow{7}{*}{$\Sigma_1$} 
& 2 &  \phantom{-}0.0383 (0.1639) & 0.8687 (0.0061) & -0.0194 (0.0301)\\ 
& 3 & \phantom{-}0.0812 (0.0550) & 0.9195 (0.0342) & \phantom{-}0.0182 (0.0210)\\ 
& 5  & \phantom{-}0.0067 (0.0295) & 0.9666 (0.0285) & \phantom{-}0.0136 (0.0194)\\ 
& 10& -0.0015 (0.0276)&0.9898 (0.0186)& \phantom{-}0.0039 (0.0120)\\ 
& 15& -0.0017 (0.0272)&0.9929 (0.0179)& \phantom{-}0.0027 (0.0110)\\ 
& 25& -0.0016 (0.0272)&0.9954 (0.0179)& \phantom{-}0.0013 (0.0102)\\ 
& \bf{Classic} &  {\bf -0.0019 (0.0262)} & {\bf 0.9986 (0.0173)} & {\bf 0.0007 (0.0084)} \\ 
\hline
\multirow{7}{*}{$\Sigma_2$} 
& 2  &  \phantom{-}0.3763 (0.1448)&0.8671 (0.0632) &-0.0163 (0.0284)\\ 
& 3  &  \phantom{-}0.0755 (0.0439)&0.9258 (0.0284)& \phantom{-}0.0151 (0.0192) \\
& 5  &  \phantom{-}0.0077 (0.0280)&0.9705 (0.0266)& \phantom{-}0.0114 (0.0182)\\ 
& 10&  \phantom{-}0.0002 (0.0267)&0.9912 (0.0182)& \phantom{-}0.0023 (0.0118)\\ 
& 15&\phantom{-} -0.0001 (0.0265)&0.9941 (0.0179)& \phantom{-}0.0021 (0.0108)\\ 
& 25& \phantom{-}0.0001 (0.0265)&0.9964 (0.0176)& \phantom{-}0.0009 (0.0100)\\ 
& \bf{Classic} & {\bf -0.0002 (0.0258)} & {\bf 0.9997 (0.0172)} & {\bf 0.0004 (0.0081)} \\ 
\hline
\multirow{7}{*}{$\Sigma_3$} 
& 2  &  \phantom{-}0.3596 (0.1310)& 0.8671 (0.0586) &-0.0150 (0.0271)\\ 
& 3  &  \phantom{-}0.0723 (0.0422)&0.9302 (0.0280)& \phantom{-}0.0113 (0.0174) \\ 
& 5  &  \phantom{-}0.0065 (0.0263)&0.9713 (0.0237)& \phantom{-}0.0102 (0.0170)\\ 
& 10&  -0.0001 (0.0252)&0.9908 (0.0174)& \phantom{-}0.0031 (0.0114)\\ 
& 15& -0.0009 (0.0249)&0.9942 (0.0170)& \phantom{-}0.0021 (0.0105)\\ 
& 25&  -0.0009 (0.0251)&0.9963 (0.0168)& \phantom{-}0.0009 (0.0096)\\ 
& \bf{Classic} & {\bf -0.0013 (0.0243)} & {\bf 0.9993 (0.0164)} & {\bf 0.0004 (0.0079)} \\ 
\hline
\multirow{7}{*}{$\Sigma_4$} 
& 2  &  \phantom{-}0.4337 (0.2211)&0.8691 (0.0847) &-0.0393 (0.0342)\\ 
& 3  & \phantom{-}0.0857 (0.0729)&0.9132 (0.0418)& \phantom{-}0.0202 (0.0250) \\ 
& 5  & \phantom{-}0.0071 (0.0355)&0.9626 (0.0366)& \phantom{-}0.0156 (0.0258)\\ 
& 10&  -0.0004 (0.0323)&0.9891 (0.0233)& \phantom{-}0.0030 (0.0172)\\ 
& 15&  -0.0009 (0.0318)&0.9930 (0.0224)& \phantom{-}0.0009 (0.0147)\\ 
& 25& -0.0010 (0.0318)&0.9953 (0.0222)& -0.0001 (0.0128)\\ 
& \bf{Classic} &  {\bf -0.0001 (0.0308)} & {\bf 0.9988 (0.0217)} & {\bf -0.0025 (0.0113)} \\ 
\hline
\multirow{7}{*}{$\Sigma_5$} 
& 2  & \phantom{-}0.3356 (0.1003)&0.8662 (0.0456) & -0.0002 (0.0093)\\ 
& 3  & \phantom{-}0.0633 (0.0246)&0.9452 (0.0184)& \phantom{-}0.0032 (0.0099) \\ 
& 5  &  \phantom{-}0.0071 (0.0157)&0.9821 (0.0140)& \phantom{-}0.0021 (0.0076)\\ 
& 10&  \phantom{-}0.0012 (0.0149)&0.9928 (0.0111)& \phantom{-}0.0009 (0.0046)\\ 
& 15& \phantom{-}0.0004 (0.0146)&0.9952 (0.0108)& \phantom{-}0.0007 (0.0038)\\ 
& 25& \phantom{-}0.0001 (0.0145)&0.9970 (0.0106)& \phantom{-}0.0003 (0.0031)\\ 
& \bf{Classic} &  {\bf -0.0004 (0.0144)} & {\bf 0.9997 (0.0104)} & {\bf 0.0000 (0.0004)} \\ 
\hline
\end{tabular}
\end{adjustbox}
 \caption{Mean (and standard errors) of the GEV parameters $(\mu, \sigma, \xi)$ of the symbolic composite MLE $\hat{\theta}^{(2)}_{SCL}$ and composite MLE $\hat{\theta}^{(2)}_{CL}$ (Classic) from $1000$ replications of the Gaussian max-stable process model, for $B\times B$ histograms for varying values of $B$.
 Results based on $N=1\,000$ observations at $K=15$ spatial locations and $T=1$ random histogram.
}
\label{tabex:table1}
\end{table}

\subsubsection{Varying the number of bins and marginal histogram dimension}

\label{ssec:62}
We generate $N=10^6$ realisations for $K=10$ locations using the covariance parameter specification $\Sigma = \Sigma_3$. Both pairwise ($B_2\times B_2$ marginal histograms) and triplewise ($B_3\times B_3\times B_3$ marginal histograms) symbolic composite MLEs, $\hat{\theta}^{(2)}_{SCL}$ and $\hat{\theta}^{(3)}_{SCL}$, were computed and compared for varying values of $B_2$ and $B_3$, constructed from a single ($T=1$) random histogram.

Table \ref{tab:table2} reports the resulting means and standard errors of  $\hat{\theta}_{SCL}^{(2)}$ and $\hat{\theta}_{SCL}^{(3)}$ obtained over 200 replicate analyses. 
Each row represents marginal pairwise and triplewise histograms with approximately equal numbers of bins (i.e.~$B_2^2\approx B_3^3$) representing approximately equivalent computational overheads.
As before, both symbolic composite MLEs converge as the number of bins increases.

When the number of bins are comparable (i.e.~$B_2^2\approx B_3^3$) the pairwise estimates invariably have smaller standard errors than the triplewise estimates. This can be attributed to the direct tradeoff between a lower resolution histogram in higher dimensions compared to a higher resolution histogram in lower dimensions, when keeping the number of histogram bins comparable. In this case, the extra lower-dimensional precision is more informative for the model parameters than higher-dimensional information, and so the pairwise estimator is more efficient.
However, when the number of bins in each margin is the same ($B_2=B_3$), so that the resolution in each dimension is the same, but where the triplewise estimator uses higher-dimensional information (using more bins), then the triplewise {\color{black}composite MLE} is naturally the most efficient.

\begin{table}[tbh]
\small
\color{black}
\centering
\begin{adjustbox}{width=0.5\textwidth}
\begin{tabular}{ |c|cc|cc|} 
\hline 
\multirow{2}{*}{$B_2^2|B_3^2$} & \multicolumn{2}{ c| }{$\sigma_{11}$}  & \multicolumn{2}{ c| }{$\sigma_{12}$}  \\
 & Pair & Triple & Pair & Triple\\
\hline
$\phantom{0}3^2 | 2^3$ & 300.62 (2.80) & 298.98 (8.45) & 150.35 (1.94) & 149.36 (5.76)  \\ 
$\phantom{0}5^2 | 3^3$&300.55 (0.95) & 300.23 (2.44) & 150.40 (0.66) & 150.09 (1.66)  \\ 
$\phantom{0}8^2 | 4^3$&300.45 (0.80) & 300.21 (1.28) & 150.31 (0.54) & 150.16 (0.86) \\ 
$11^2 | 5^3 $&300.57 (0.72) & 300.42 (0.91) & 150.39 (0.46) & 150.30 (0.62) \\ 
\hline
\multicolumn{5}{ c }{} \\
\hline
\multirow{2}{*}{$B_2^2|B_3^2$} & \multicolumn{2}{ c| }{$\sigma_{22}$}  & \multicolumn{2}{ c| }{$\mu$}  \\
 & Pair & Triple & Pair & Triple \\
\hline
$\phantom{0}3^2 | 2^3$&200.14 (1.74) & 199.68 (5.46)& 0.0426 (0.0217) & 0.1515 (0.0494)\\
$\phantom{0}5^2 | 3^3$&200.26 (0.55) & 200.02 (1.50)& 0.0016 (0.0025) & 0.0411 (0.0209)\\
$\phantom{0}8^2 | 4^3$ &200.20 (0.50) & 200.07 (0.82)& 0.0001 (0.0007) & 0.0093 (0.0079)\\
$11^2 | 5^3$&200.22 (0.38) & 200.19 (0.56)&0.0000 (0.0008) & 0.0015 (0.0023) \\
\hline
\multicolumn{5}{ c }{} \\
\hline
\multirow{2}{*}{$B_2^2|B_3^2$} & \multicolumn{2}{ c| }{$\sigma$}  & \multicolumn{2}{ c| }{$\xi$} \\
 & Pair & Triple & Pair & Triple \\
\hline
$\phantom{0}3^2 | 2^3$&0.9803 (0.0094)&0.9718 (0.0112)&0.0039 (0.0023) & 0.0004 (0.0055)\\
$\phantom{0}5^2 | 3^3$&0.9978 (0.0033)&0.9807 (0.0092)&0.0008 (0.0013) & 0.0037 (0.0023)\\
$\phantom{0}8^2 | 4^3$&0.9999 (0.0007)&0.9926 (0.0056)&0.0001 (0.0001) & 0.0020 (0.0016)\\
$11^2 | 5^3$&0.9999 (0.0001)&0.9978 (0.0029)&0.0000 (0.0001) & 0.0008 (0.0011)\\
\hline
\end{tabular}
}
\end{adjustbox}
 \caption{Mean (and standard errors) of the pairwise ($\hat{\theta}^{(2)}_{SCL}$) and triplewise ($\hat{\theta}^{(3)}_{SCL}$) symbolic composite MLEs from $200$ replications of the Gaussian max-stable process model for $B_2\times B_2$ (pairwise) and $B_3\times B_3\times B_3$ (triplewise) histograms, with varying $B_2$, $B_3$.
Rows correspond to $B^2_2\approx B^3_3$ to compare approximately equal numbers of histogram bins.
Results based on $N=10^6$ observations at $K=10$ spatial locations, $T=1$ random histogram and $\Sigma=\Sigma_3$.
}
\label{tab:table2}
\end{table}

\subsubsection{Varying the number of spatial locations, $K$}
\label{ssec:63}

We generate $N=10^6$ realisations at $K$ locations (for varying $K$) using the covariance parameter specification $\Sigma=\Sigma_3$. The random locations for smaller $K$ are a subset of those for larger $K$.
Both pairwise and triplewise symbolic composite MLEs, $\hat{\theta}_{SCL}^{(2)}$ and $\hat{\theta}_{SCL}^{(3)}$, are computed, using $B_2\times B_2$ and $B_3\times B_3\times B_3$ random marginal histograms, where $B_2=8$ and $B_3=4$ so that each marginal histogram has 64 bins.

Table \ref{tab:table3} reports the resulting means and standard errors of $\hat{\theta}_{SCL}^{(2)}$ and $\hat{\theta}_{SCL}^{(3)}$ for different values of $K$, based on 200 replicate analyses.
As expected, as $K$ increases both {\color{black}composite MLEs} become increasingly accurate, particularly the dependence parameters ($\sigma_{11}, \sigma_{12}, \sigma_{22}$), as the amount of spatial information increases, with the pairwise {\color{black}composite MLEs} producing more accurate estimates for an equivalent number of bins.
These results are consistent with those for standard pairwise and triplewise composite MLEs seen in e.g.~\cite{padoan+rs10} and \cite{Genton2011}.

\begin{table}[tbh]
\small
\centering
\color{black}
\begin{adjustbox}{width=0.5\textwidth}
\begin{tabular}{ |c|cc|cc|} 
\hline 
\multirow{2}{*}{$K$} & \multicolumn{2}{ c| }{$\sigma_{11}$}  & \multicolumn{2}{ c| }{$\sigma_{12}$}   \\
& Pair & Triple & Pair & Triple \\
\hline
$3$ & 300.44 (5.80) & 299.24 (13.37) & 150.30 (2.41) & 150.02 (6.75)  \\ 
$5$&300.35 (1.53) & 299.95 (\phantom{-}2.37) & 150.28 (1.10) & 150.02 (1.99)  \\ 
$10$&300.21 (0.88) & 299.95 (\phantom{-}1.22) & 150.15 (0.59) & 149.99 (0.83) \\ 
$15$&300.19 (0.71) & 299.93 (\phantom{-}1.12) & 150.12 (0.48) & 150.00 (0.73)  \\ 
$20$&300.20 (0.78) & 299.99 (\phantom{-}0.99) & 150.14 (0.47) & 150.02 (0.70) \\ 
\hline
\multicolumn{5}{ c }{} \\
\hline
\multirow{2}{*}{$K$}  & \multicolumn{2}{ c| }{$\sigma_{22}$}& \multicolumn{2}{ c| }{$\mu$}   \\
& Pair & Triple & Pair & Triple \\
\hline
$3$&201.55 (11.12) & 200.12 (7.84)& -0.00003 (0.0011) & 0.00727 (0.0102)\\
$5$&200.22 (\phantom{-}1.00) & 199.98 (1.89) & -0.00002 (0.0010) & 0.00671 (0.0088)\\
$10$ &200.10 (\phantom{-}0.53) & 199.94 (0.77)& -0.00006 (0.0009) & 0.00595 (0.0068)\\
$15$&200.06 (\phantom{-}0.46) & 200.00 (0.72)&-0.00004 (0.0001) & 0.00553 (0.0054)\\
$20$&200.08 (\phantom{-}0.44) & 199.99(\phantom{-}0.69) &-0.00005 (0.0001) & 0.00524 (0.0053) \\
\hline
\multicolumn{5}{ c }{} \\
\hline
\multirow{2}{*}{$K$}  & \multicolumn{2}{ c| }{$\sigma$}  & \multicolumn{2}{ c| }{$\xi$}\\
& Pair & Triple & Pair & Triple \\
\hline
$3$&0.9999 (0.0009)&0.9947 (0.0069)&0.00006 (0.00062) & 0.00121 (0.00193)\\
$5$&0.9999 (0.0008)&0.9950 (0.0064)&0.00008 (0.00059) & 0.00111 (0.00187)\\
$10$&0.9999 (0.0007)&0.9956 (0.0047)&0.00009 (0.00048) & 0.00093 (0.00133)\\
$15$&0.9999 (0.0007)&0.9958 (0.0042)&0.00007 (0.00042) & 0.00092 (0.00131)\\
$20$&0.9999 (0.0007)&0.9961 (0.0039)&0.00006 (0.00048) & 0.00080 (0.00121)\\
\hline

\end{tabular}
\end{adjustbox}
 \caption{
 Mean (and standard errors) of the pairwise ($\hat{\theta}^{(2)}_{SCL}$) and triplewise ($\hat{\theta}^{(3)}_{SCL}$) symbolic composite MLEs from $200$ replications of the Gaussian max-stable process model for $B_2\times B_2$ (pairwise) and $B_3\times B_3\times B_3$ (triplewise) histograms, with varying $K$.
Results based on $N=10^6$ observations in $T=1$ random histogram with $B_2=8$ and $B_3=4$ (so that $B^2_2=B_3^3$) and $\Sigma=\Sigma_3$.
 }
\label{tab:table3}
\end{table}

\subsubsection{Varying the number of underlying observations, $N$}
\label{ssec:64}

One of the motivations for aggregating micro-data into random histograms before an analysis is that the analysis, while losing some information in the data, will be much faster.
We generate $N=10^3,\ldots,10^7$ realisations for $K=10$ locations using the covariance parameter specification $\Sigma = \Sigma_3$. We compute standard pairwise composite ($\hat \theta_{CL}^{(2)}$) and symbolic pairwise composite  ($\hat \theta_{SCL}^{(2)}$) MLEs,
with $B_2=25$ and $T=1$.

Table \ref{tab:table4} reports the resulting means and standard errors of $\hat{\theta}_{CL}^{(2)}$ and $\hat{\theta}_{SCL}^{(2)}$ for different values of $N$, based on 100 replicate analyses. 
As expected, as $N$ increases the {\color{black}composite MLEs} become increasingly accurate, with the standard composite MLEs outperforming the symbolic composite MLEs, although the difference here is relatively minor as we are using $25\times 25$ histogram bins in each pairwise comparison.
However, it was not computationally viable to compute $\hat{\theta}^{(2)}_{CL}$ for $N\geq 10^6$.
To explore this in more detail, these simulations were repeated for
$K=20,50,100$ spatial locations, and a slightly smaller range of observed data ($N=1\,000$ to $500\,000$) to provide a better comparison with the standard composite MLEs.

Table \ref{tab:tabletimes} summarises the mean computation times (in seconds) for different stages involved in computing the {\color{black}composite MLEs}, based on 10 replicate analyses.
Simply in terms of optimising the respective likelihood functions, the symbolic composite likelihood ($t_s$) is much more efficient than the equivalent composite likelihood ($t_c$). The computational overheads of the former are essentially constant with respect to $N$, and so these are largely driven by the number of pairwise components ($K/(K-1)/2$) in the likelihood. The computational overheads of the composite likelihood are driven both by $N$ and $K$, and so computing $\hat{\theta}^{(2)}_{CL}$ becomes largely impractical when either becomes moderately large. Clearly computation of $\hat{\theta}^{(2)}_{SCL}$ would take similar times to those in Table \ref{tab:tabletimes} for considerably larger $N$. 

An additional step in computing $\hat{\theta}^{(2)}_{SCL}$ is construction of all bivariate marginal histograms $\bS^\bi, \bi\in{\mathcal I}_2$. We constructed these in two alternative ways: using the {\tt R} function {\tt hist} ($t_{hist R}$) and the {\tt R} package {\tt DeltaRho}  ($t_{hist DR}$) which provides an interface to {\tt map-reduce} functionality whereby the histograms can be constructed in parallel on multiple processors and machines, and then combined. 

For small values of $N$, using the simple {\tt hist} function on a local machine is quicker than using {\tt DeltaRho} and communicating between multiple machines. However, {\tt DeltaRho} increasingly outperforms {\tt hist} as the number of datapoints $N$ increases. Our {\tt DeltaRho} setup was modest with only 4 parallel machines; more expansive setups could drastically reduce histogram construction time for large $N$.
Regardless of the histogram construction method adopted, it is clear that computing the symbolic composite MLE is considerably more efficient than the standard composite MLE.

\begin{table}[h]
\color{black}
\small
\centering
\begin{adjustbox}{width=0.6\textwidth}
\small
\begin{tabular}{ |c|cc|cc|} 
\hline 
\multirow{2}{*}{$N$}  & \multicolumn{2}{ c| }{$\sigma_{11}$}  & \multicolumn{2}{ c| }{$\sigma_{12}$} \\
 & Classic & Pair & Classic & Pair \\
\hline
$10^3$&299.48 (17.09) & 298.11 (17.24) & 149.90 (10.37) & 148.84 (11.05)  \\ 
$10^4$&299.07 (\phantom{-}5.76) & 298.56 (\phantom{-}6.07) & 149.65 (\phantom{-}3.26) & 149.09 (\phantom{-}3.63) \\ 
$10^5$&300.56 (\phantom{-}1.56) & 300.49 (\phantom{-}2.07) & 150.42 (\phantom{-}0.98) & 150.32 (\phantom{-}1.27)  \\ 
$10^6$&--                    & 300.21 (\phantom{-}0.61) &--                     & 150.18 (\phantom{-}0.45)  \\ 
$10^7$&--                     & 300.13 (\phantom{-}0.23) &--                     & 150.06 (\phantom{-}0.17)  \\ 
\hline
\multicolumn{5}{ c }{} \\
\hline
\multirow{2}{*}{$N$} & \multicolumn{2}{ c| }{$\sigma_{22}$}  & \multicolumn{2}{ c| }{$\mu$} \\
& Classic & Pair & Classic & Pair \\
\hline
$10^3$&200.45 (11.05) & 200.11 (11.69)& -0.0074 (0.0280) & -0.0077 (0.0286) \\ 
$10^4$ &199.92 (\phantom{-}3.32) & 199.39 (\phantom{-}3.70)& -0.0017 (0.0074) & -0.0013 (0.0076) \\ 
$10^5$&200.28 (\phantom{-}1.14) & 200.18 (\phantom{-}1.49)& -0.0002 (0.0021) & -0.0002 (0.0025) \\ 
$10^6$ &--                     & 200.14 (\phantom{-}0.43) &--                            & \phantom{-}0.0000 (0.0007)\\ 
$10^7$&--                     & 200.02 (\phantom{-}0.18) &--                            & -0.0001 (0.0002)\\ 
\hline
\multicolumn{5}{ c }{} \\
\hline
\multirow{2}{*}{$N$} & \multicolumn{2}{ c| }{$\sigma$}  & \multicolumn{2}{ c| }{$\xi$} \\
 & Classic & Pair & Classic \\
\hline
$10^3$&0.9972 (0.0169)&0.9964 (0.0170)& \phantom{-}0.0016 (0.0115) & \phantom{-}0.0024 (0.0123)\\
$10^4$&0.9989 (0.0051)&0.9988 (0.0052)&-0.0002 (0.0039) & -0.0002 (0.0040)\\
$10^5$&1.0000 (0.0014)&1.0000 (0.0015)& \phantom{-}0.0001 (0.0010) & \phantom{-}0.0001 (0.0013)\\
$10^6$&--                          &1.0000 (0.0004)&--                            & \phantom{-}0.0000 (0.0004)\\
$10^7$&--                         &1.0000 (0.0001)&--                            & \phantom{-}0.0000 (0.0001)\\
\hline

\end{tabular}
\end{adjustbox}
 \caption{Mean (and standard errors) of the standard pairwise composite ($\hat{\theta}^{(2)}_{CL}$) and symbolic pairwise composite ($\hat{\theta}^{(2)}_{SCL}$) MLEs from $100$ replications of the Gaussian max-stable process model with $B_2\times B_2$ histograms with $B_2=25$.
Results are based on $K=10$ spatial locations, $T=1$ random histogram and $\Sigma=\Sigma_3$.
 }
\label{tab:table4}
\end{table}

\begin{table}[h]
\centering
\begin{adjustbox}{width=0.6\textwidth}
\begin{tabular}{ |c|cccc|cccc|} 
\hline
\multirow{2}{*}{$N$} & \multicolumn{4}{ c| }{$K=10$}& \multicolumn{4}{ c| }{$K=20$} \\
 &$t_c$&$t_{s}$ & $t_{hist DR}$ & $t_{hist R}$   & $t_c$  &$t_{s}$ & $t_{hist DR}$ & $t_{hist R}$ \\
\hline
$1\,000$& \phantom{00\,0}71.9 &22.5   & 0.8  &0.1  & \phantom{00\,}383.4 & \phantom{0}79.6 & \phantom{0}1.8 & \phantom{0}0.4\\ 
$5\,000$& \phantom{00\,}291.8&19.0   & 0.8  &0.3    & \phantom{0}1\,578.2 & \phantom{0}99.3 & \phantom{0}2.1& \phantom{0}1.0\\ 
$10\,000$  & \phantom{00\,}591.7& 23.8&0.9 & 0.5& \phantom{0}3\,125.4 & 103.2 & \phantom{0}2.4 & \phantom{0}1.8\\ 
$50\,000$& \phantom{0}2\,626.8& 24.2 &1.7 & 2.1& 20\,459.4 &107.3 & \phantom{0}4.5 & \phantom{0}7.6 \\ 
$100\,000$&\phantom{0}5\,610.7& 25.4  &2.4 & 4.2&--     &115.0     & \phantom{0}6.9 & 14.9 \\ 
$500\,000$ &31\,083.1 & 23.2  &7.5 & 20.6&--     & \phantom{0}96.1     & 26.6 & 73.5\\ 
\hline
\multicolumn{9}{ c }{} \\
\hline
\multirow{2}{*}{$N$} &  \multicolumn{4}{ c| }{$K=50$}&\multicolumn{4}{ c| }{$K=100$}  \\
 &$t_c$  &$t_{s}$ & $t_{hist DR}$ & $t_{hist R}$ & $t_c$ &$t_{s}$ & $t_{histDR}$ & $t_{hist R}$ \\
\hline
$1\,000$& \phantom{0}7\,333.9&528.5 & \phantom{00}9.3 & \phantom{00}3.0 & --&2\,238.0 & \phantom{0}78.8 & \phantom{00}12.0 \\ 
$5\,000$& 27\,616.5&665.1& \phantom{0}10.6 & \phantom{00}7.7   & --&2\,650.2 & \phantom{0}81.7 & \phantom{00}30.9 \\ 
$10\,000$  &--   & 696.3   &     \phantom{0}12.4 & \phantom{0}13.5   &--&2\,356.6 & \phantom{0}85.8 & \phantom{00}54.1\\ 
$50\,000$&--   & 744.8    &    \phantom{0} 24.8 & \phantom{0}59.0   &--&2\,300.6 & 131.6 & \phantom{0}237.0 \\ 
$100\,000$&--     & 768.1   &     \phantom{0}41.3 & 115.7  & --&2\,766.9 & 188.2 & \phantom{0}461.8 \\ 
$500\,000$ &--    & 802.9   &    156.1 & 561.3   &--&3\,111.5 & 627.1& 2\,243.5 \\ 
\hline

\end{tabular}
\end{adjustbox}
 \caption{Mean computation times (seconds) for different components involved in computing $\hat{\theta}^{(2)}_{CL}$ and $\hat{\theta}^{(2)}_{SCL}$ for different classical dataset sizes $N$ and number of spatial locations $K$, based on $10$ replicate analyses.  
 Columns $t_c$ and $t_s$ respectively show the time taken to optimise the standard composite and symbolic composite likelihood functions. Columns $t_{hist DR}$ and $t_{hist R}$ show the time taken to aggregate the data into histograms using {\tt DeltaRho} and {\tt R} function {\tt hist} respectively.
Results are based on $T=1$ random histogram and $\Sigma=\Sigma_3$.
 }
\label{tab:tabletimes}
\end{table}

\subsubsection{Varying the number of histograms, $T$}
\label{ssec:65}

Until now the $N$ observed datapoints have been aggregated into a single histogram, $T=1$ (or more precisely one low-dimensional marginal histogram per composite likelihood component).
If each histogram $\bS_1,\ldots,\bS_T$ has exactly the same bins then collapsing these to a single histogram, as discussed in Section \ref{ssec:a2}, will produce the same symbolic composite MLE as if $T>1$ histograms were used. 
However the number of random histograms $T$ will affect the standard errors of $\hat{\theta}^{(j)}_{SCL}$, as discussed in Section \ref{ssec:999}. That is, by aggregating the spatially observed micro-data over multiple time points, there is a loss of information in knowing which observations at location $t_i$ occurred at the same time as observations at location $t_j$ within the same random histogram. This results in a loss of spatial information, which will impact the efficiency of the symbolic likelihood estimators.

\begin{figure}[h]
\centering
\includegraphics[page=1,width=0.6\textwidth]{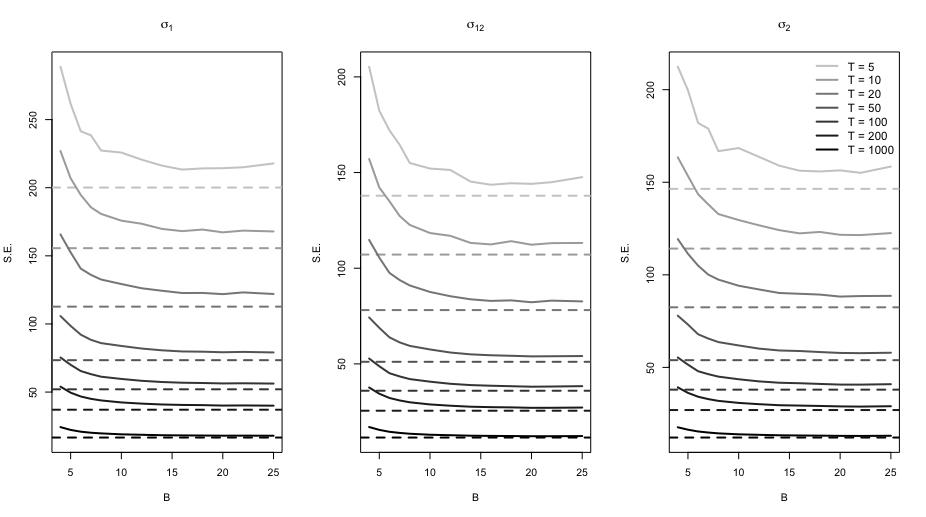}
 \caption{Godambe standard errors (solid lines) for the dependance parameters ($\sigma_{11}, \sigma_{12}, \sigma_{22}$) of $\hat{\theta}^{(2)}_{SCL}$ for varying number of random histograms $T$,
 and number of marginal histogram bins $B^2$. 
 Dashed horizontal lines denote the appropriate term of the limit \eqref{classicgodambe} of the variability matrix $\hat{J}(\hat{\theta}^{(2)}_{SCL})$.
 Results are based on $N=1\,000$ observations with $\Sigma=\Sigma_3$.
 }
\label{fig:Tplot}
\end{figure}

To examine this we generate $N=1\,000$ realisations for $K=10$ spatial locations using the covariance parameter specification $\Sigma=\Sigma_3$. We compute the standard composite ($\hat{\theta}^{(2)}_{CL})$ and symbolic ($\hat{\theta}^{(2)}_{SCL}$) pairwise {\color{black}composite MLEs} when aggregating the observations equally into $T=4,5,10,20,40,50,100,200$ and $1000$ histograms $\bS_t$ (so that for $T=1\,000$ we have 1 observation per random histogram), with  $B\times B=25^2$ bins in each pairwise marginal histogram.
The means of the Godambe standard errors for the {\color{black}composite MLEs} for each value of $T$ are reported in Table \ref{tab:table5}, based on 1\,000 replicate analyses. 
This procedure is then repeated $100$ times while varying the number of marginal histogram bins ($B^2$), with the results illustrated in Figure \ref{fig:Tplot}.

From Table \ref{tab:table5}, for a small number of histograms the estimated standard errors are large compared to the standard composite likelihood estimates due to the significant loss of temporal information. 
As $T$ increases these standard errors reduce as more temporal information is recovered. With $T=N$ (and one data point per histogram) the standard errors become comparable, although the location of the single datapoint within each histogram for $T=N$ is still uncertain, and so unless the number of bins also increases, the standard errors of the {\color{black}symbolic composite MLE} will be larger than those of the standard composite MLE, even for $T=N$.
Figure \ref{fig:Tplot} illustrates how the mean Godambe standard errors, for fixed $T$, approach the (square root of the) appropriate diagonal term of the limit \eqref{classicgodambe} of the variability matrix $\hat{J}(\hat{\theta}^{(2)}_{SCL})$, as the number of histogram bins becomes large.
As $T\rightarrow N$ this limit (horizontal dashed lines) approaches the equivalent standard errors under the standard composite likelihood (the lowest horizontal dashed line).

Of course, while standard error accuracy increases for larger $T$, computational overheads increase in  proportion to $T$.
Hence in practice, and with equal bins over all histograms, to compute the symbolic composite MLE $\hat{\theta}^{(j)}_{SCL}$ we would use $T=1$, whereas to compute standard errors we would use as small a number of histograms as possible (to maximise computational efficiency) such that the scale of the standard errors is acceptable within the context of the given analysis.

\begin{table}[h]
\centering
\begin{adjustbox}{width=0.6\textwidth}
\begin{tabular}{ |c|cccccc|} 
\hline 
 $T$& $\sigma_{11}$ & $\sigma_{12}$ & $\sigma_{22}$ & $\mu$ & $\sigma$ & $\xi$\\
\hline
5 & 217.81 & 147.60 & 158.48 & 0.31 & 0.19 & 0.13 \\ 
  10 & 167.90 & 113.21 & 122.55 & 0.23 & 0.15 & 0.10 \\ 
  20 & 122.00 &  \phantom{0}82.66 & \phantom{0}88.64 & 0.17 & 0.11 & 0.07 \\ 
  50 & \phantom{0}79.09 & \phantom{0}54.10 &  \phantom{0}57.91 & 0.11 & 0.07 & 0.05 \\ 
  100 & \phantom{0}56.23 & \phantom{0}38.37 &  \phantom{0}40.93 & 0.08 & 0.05 & 0.03 \\ 
  200 & \phantom{0}40.01 & \phantom{0}27.19 &  \phantom{0}29.02 & 0.06 & 0.04 & 0.02 \\ 
  1000 & \phantom{0}17.94 & \phantom{0}12.28 &  \phantom{0}13.07 & 0.03 & 0.02 & 0.01 \\ 
\bf{Classic} & {\bf 16.65} & {\bf 11.53} & {\bf 12.69} &  {\bf 0.021} & {\bf 0.014} &  {\bf 0.008} \\ 
\hline
\end{tabular}
\end{adjustbox}
 \caption{Means of the estimated Godambe standard errors of $\hat{\theta}^{(2)}_{SCL}$ and $\hat{\theta}^{(2)}_{CL}$ for different numbers of random histograms, $T$, based on $1\,000$ replicate analyses.
 Results are based on $N=1\,000$ observations with $B=25$ and $\Sigma=\Sigma_3$.
 }
\label{tab:table5}
\end{table}

%
%
%
\section{Analysis of millennial scale climate extremes}
\label{sec:6}
\begin{figure}[h]
\centering
\includegraphics[page=1,scale=0.26]{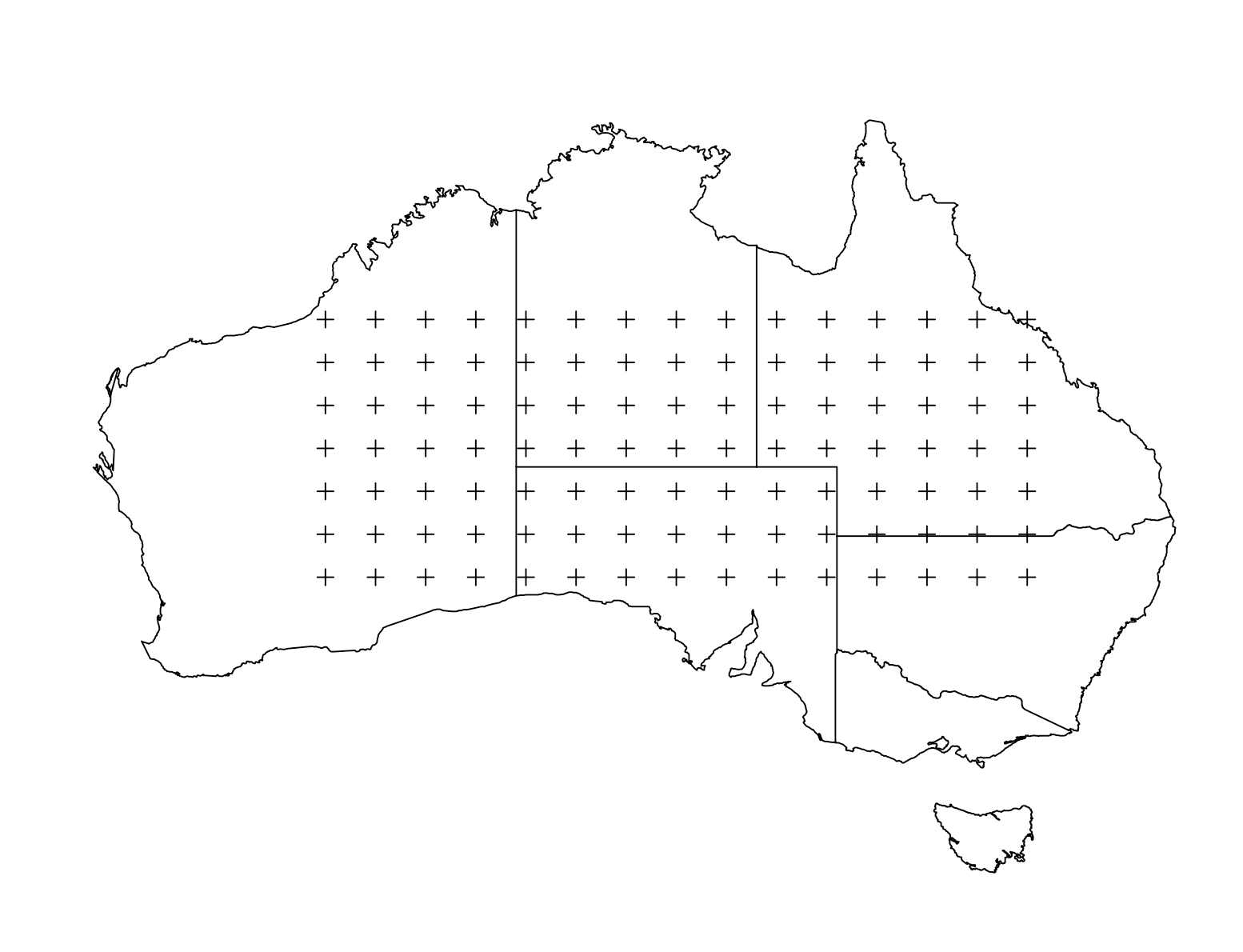}
 \caption{$K=105$ spatial locations for the historical and future-simulated temperature data over Australia. Each cross represents the midpoint of a $1.875^{\circ} \times 1.875^{\circ}$ box in a spatial grid.}
\label{fig:aus}
\end{figure}

We consider daily maxima of historical temperature data (1850--2006) and future simulated temperature data (2006--2100) simulated using the CSIRO Mk3.6 climate model, for 105 grid locations (considered as the spatial co-ordinates) at the centre of  $1.875^{\circ} \times 1.875^{\circ}$ grid cells over Australia (Figure \ref{fig:aus}).
Two different scenarios (RCP4.5 and RCP8.5) are used to generate the future data, which represent two of the four greenhouse gas scenarios projected by the Intergovernmental Panel on Climate Change (IPCC) based on how much greenhouse gases are emitted in future years \citep{IPCCreport}. 
Due to seasonal periodicity, only data from $90$ days across the summer months (December--February) are considered, to induce approximate stationarity of the process. 
Due to the temporal dependence evident in the RCP4.5 and RCP8.5 data the daily maximum temperatures at each spatial location were linearly detrended, so that the resulting block-maxima constitute the largest deviation above the mean temperature.
Maxima  are computed over 15-day blocks, resulting in 6 observations per year, and $N=936$ and $570$ total observations per location for the historical and climate model data respectively. 
\textcolor{black}{For this block size, we examined qq-plots of empirical versus theoretical GEV quantiles for each spatial location and each data scenario (historical or future).
The fidelity of the GEV limit was very good for both future scenarios, and less good, but reasonable, for the historical data (results not shown). }

%
Following \citet{padoan+rs10} and \citet{Blanchet2011} we fit the Gaussian max-stable process \citep{Smith1990} model with spatially varying marginal parameters, in particular with
\begin{align*}
	\mu (k) &= \alpha_{0} + \alpha_{1}x(k)+ \alpha_{2}y(k),\\
	  \sigma (k)& = \beta_{0} + \beta_{1}x(k)+ \beta_{2}y(k),\\
	   \xi (k)& = \xi,
\end{align*}
where $(x(k),y(k))$ are the spatial co-ordinates of the $k$-th location. Other co-variates (such as altitude) were not considered due to the reasonably flat nature of the topography across the study region.

{\color{black}
For this dataset, the computation times presented in Table \ref{tab:tabletimes} (with $K=100$ and $N=1\,000$) indicate that the symbolic composite pairwise approach can be considerably more efficient than
the classical composite likelihood function. Further, 
}Table \ref{tab:terms} lists the total number of terms in the standard pairwise composite likelihood, $\ell_{CL}^{(2)}(\theta)$,  and the symbolic composite likelihood, $\ell_{CL}^{(2)}(\theta)$, for a single ($T=1$) bivariate $B\times B$ histogram with $B=15, 20, 25, 30$. While the number of terms in the symbolic likelihood is guaranteed to be lower than the standard likelihood if $B^2<N$, in practice the number of non-empty histogram bins contributing to the likelihood (centre column, Table \ref{tab:terms}) {\color{black}can be considerably smaller}, particularly for strongly dependent data. For the current analyses, the symbolic composite likelihood has significantly fewer terms, leading to substantially faster optimisation and lower computational costs than the standard composite likelihood.
As discussed in Section \ref{sec:compsym}, the symbolic composite MLE ($\hat{\theta}^{(2)}_{SCL}$) can be computed exactly with $T=1$ random histogram, and so this optimisation (which evaluates the target function many times) can be very efficient. In contrast, $T=N$ histograms are required for the best variance estimates (see Table \ref{tab:table5}), and so the resulting  computational overheads are comparable to that of the standard composite likelihood (though these are only a small proportion of total computation).

\begin{table}[h] 
\centering
\begin{adjustbox}{width=0.6\textwidth}
\small
\begin{tabular}{ |c|ccc|} 
\hline 
 &Historical &Actual RCP4.5/8.5  & Maximum RCP4.5/8.5\\
 $B$& ($N=936$) & ($N=570$)  & ($N=570$)\\
\hline
$15$ &\phantom{1}\,642\,898&\phantom{1}\,529\,584&1\,228\,500 \\ 
$20$&\phantom{1}\,960\,403&\phantom{1}\,774\,060&2\,184\,000\\ 
$25$&1\,286\,714&1\,016\,565&3\,412\,500\\ 
$30$&1\,609\,923&1\,247\,465&4\,914\,000\\
\bf{Classic} & {\bf 5\,110\,560} & {\bf 3\,112\,200} & {\bf 3\,112\,200} \\
\hline

\end{tabular}
\end{adjustbox}
 \caption{Total number of terms in each pairwise composite likelihood function for $N=936, 570$ block maxima over $K=105$ spatial locations. For standard composite likelihoods this corresponds to $NK(K-1)/2$ terms. For the symbolic composite likelihood constructed using a single ($T=1$) $B\times B$ histogram, this corresponds to a maximum of $B^2K(K-1)/2$ terms.
The actual number of symbolic composite likelihood terms corresponds to the number of non-empty histogram bins. 
 }
\label{tab:terms}
\end{table}

Table \ref{tab:tablereal} displays the symbolic composite MLEs (and standard errors) of the three dependence parameters and the marginal shape parameter $\xi$ for the Smith model, calculated using $B=15, 20, 25, 30$.
Comparable {\color{black}estimates} are obtained for each value of $B$, with some clear convergence in both the point estimates and their standard errors as the resolution of each histogram increases. While the standard errors are naturally larger than those under the standard composite likelihood by construction, they are sufficiently small compared to the magnitude of the {\color{black}composite MLE} in order to make meaningful inference.

\begin{table}[h]
\centering
\begin{adjustbox}{width=0.6\textwidth}
\small
\begin{tabular}{ |c|cccc|} 
\multicolumn{5}{ c }{Historical Data}\\
\hline
 $B$ & $\sigma_{11}$  & $\sigma_{12}$  & $\sigma_{22}$ & $\xi$\\
\hline
$15$ &176.4 (2.85)&-28.7 (0.32)&76.8 (3.29) &-0.266 (0.053) \\ 
$20$&164.2 (2.89)&-29.3 (0.30) &74.3 (4.69)&-0.264 (0.049)\\ 
$25$&162.4 (2.17)&-29.9 (0.33)&75.3 (2.84)&-0.264 (0.049)\\  
$30$&161.6 (2.01)&-32.3 (0.29)&74.4 (2.34)&-0.264 (0.050)\\  
\hline
\multicolumn{5}{ c }{}\\
\multicolumn{5}{ c }{RCP4.5 Data}\\
\hline
 $B$ & $\sigma_{11}$  & $\sigma_{12}$  & $\sigma_{22}$ & $\xi$\\
\hline
$15$ &160.9 (9.42)&-34.1 (0.83)&79.0 (2.22)&-0.249 (0.074) \\ 
$20$&163.5 (5.95)&-41.1 (0.73)&77.6 (2.45)&-0.249 (0.076)\\ 
$25$&150.3 (3.49)&-33.1 (0.65)&70.7 (1.70)&-0.250 (0.073)\\ 
$30$&150.2 (1.50)&-31.6 (0.24)&70.7 (1.54)&-0.250 (0.069)\\ 
\hline
\multicolumn{5}{ c }{}\\
\multicolumn{5}{ c }{RCP8.5 Data}\\
\hline
 $B$ & $\sigma_{11}$  & $\sigma_{12}$  & $\sigma_{22}$ & $\xi$\\
\hline
$15$ &128.7 (8.60) &-19.6 (0.92)&67.7 (3.92)&-0.232 (0.061) \\ 
$20$&128.0 (6.30)&-19.6 (1.29)&66.6 (3.32)&-0.231 (0.059)\\ 
$25$&136.0 (3.95)&-15.1 (0.93)&59.4 (3.17)&-0.234 (0.060)\\ 
$30$&129.9 (4.01)&-13.6 (0.83)&56.4 (2.94)&-0.233 (0.055)\\ 
\hline

\end{tabular}
\end{adjustbox}
 \caption{The mean and standard errors of the {\color{black}composite MLEs} for $\Sigma$ obtained for the $105$ locations across Australia from the bivariate symbolic composite log-likelihood function for $B=15,20,25,30$.}
\label{tab:tablereal}
\end{table}

Compared to the observed historical extremes, 
we can see a slight increase in spatial dependence for the RCP4.5 scenario data and a significant decrease in dependence for the RCP8.5 scenario. 

The marginal shape parameter $\xi$ is negative for all three datasets, with larger {\color{black}composite MLEs} estimated for the future-simulated data compared to the historical data. This implies that the RCP4.5 and RCP8.5 data have higher upper bounds than that of the historical dataset, meaning larger deviations from the mean are expected for the future scenarios.

\begin{figure}[t!]
\centering
\includegraphics[page=1,scale=0.132]{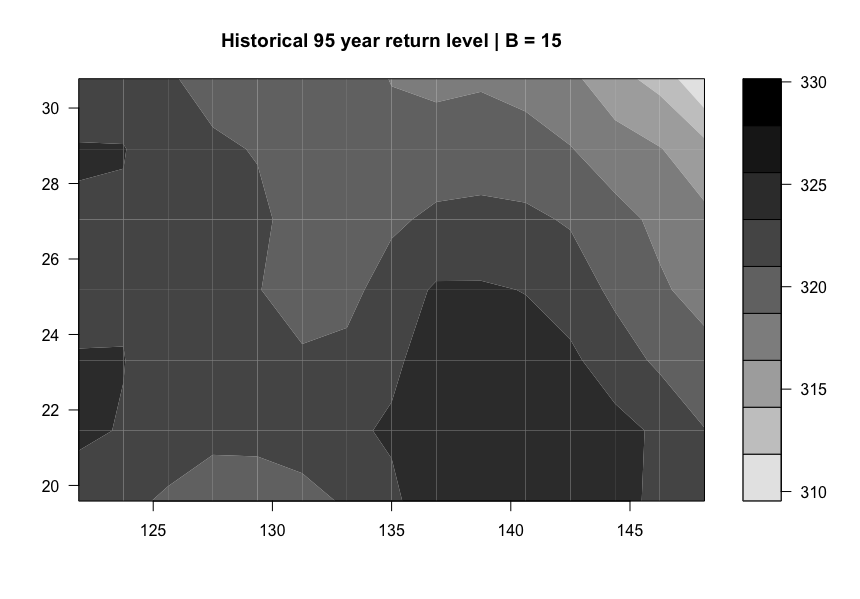}
\includegraphics[page=1,scale=0.132]{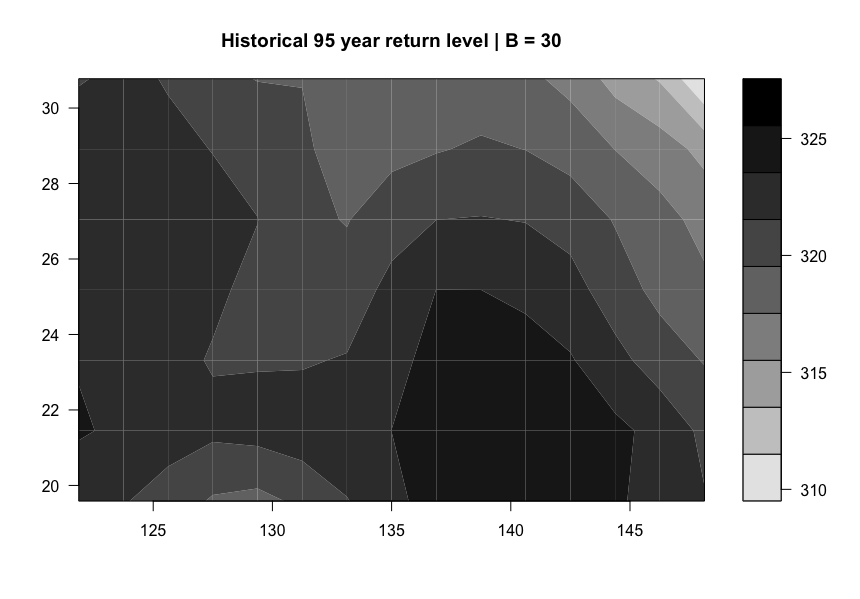}
\includegraphics[page=1,scale=0.132]{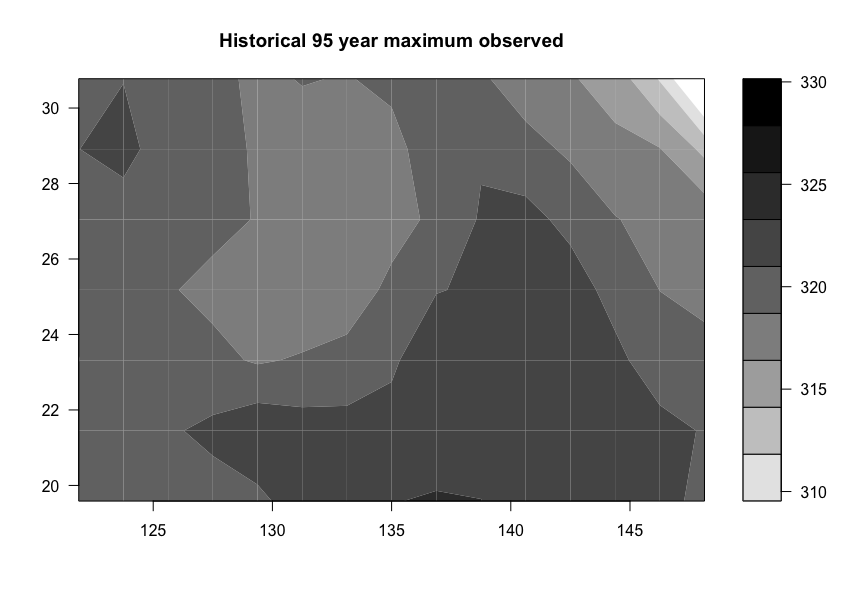}
\includegraphics[page=1,scale=0.132]{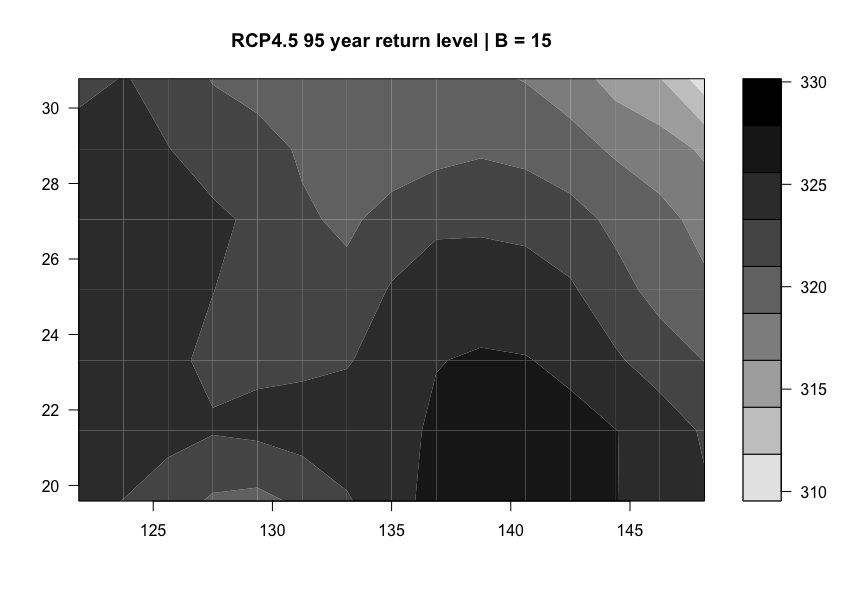}
\includegraphics[page=1,scale=0.132]{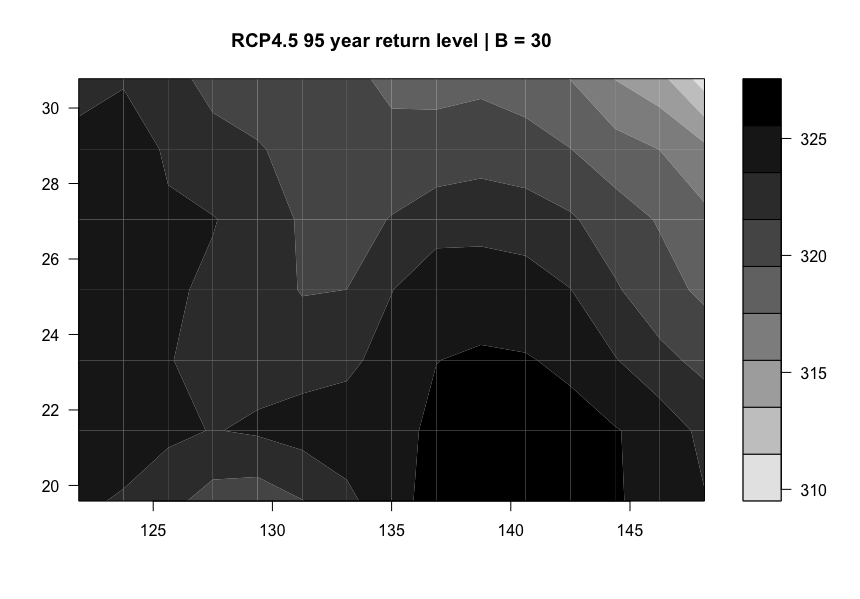}
\includegraphics[page=1,scale=0.132]{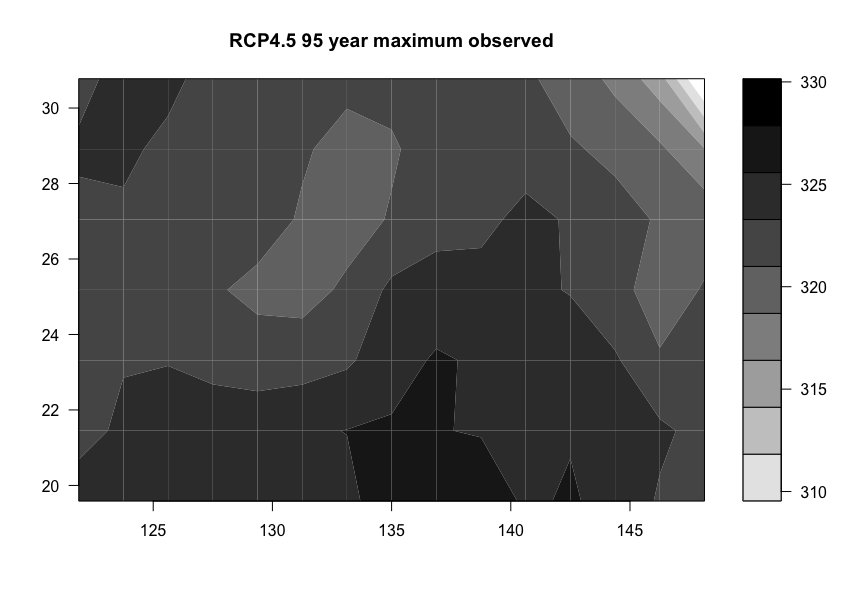}
\includegraphics[page=1,scale=0.132]{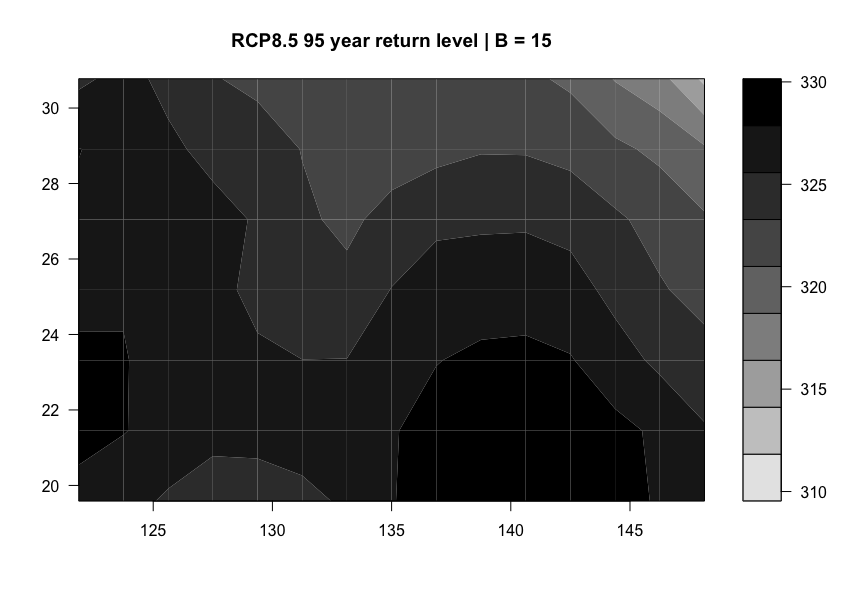}
\includegraphics[page=1,scale=0.132]{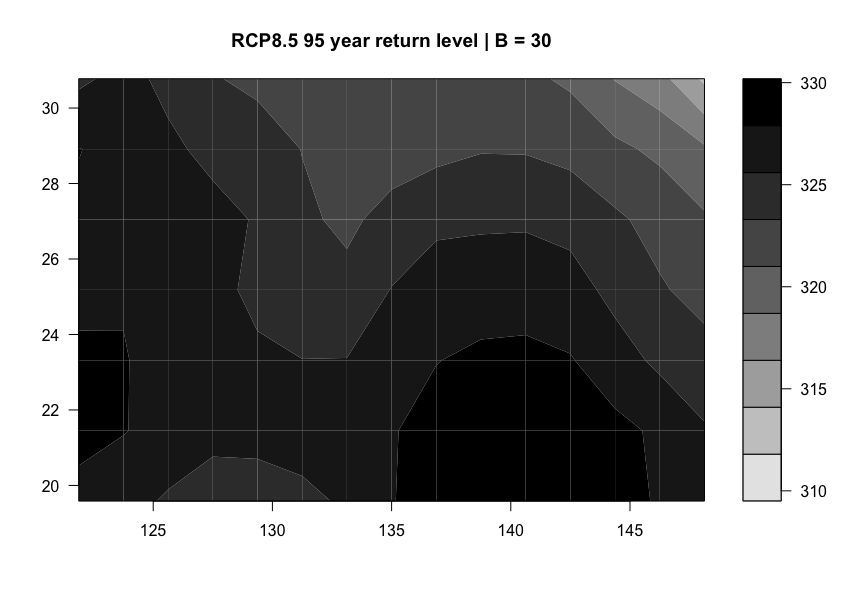}
\includegraphics[page=1,scale=0.132]{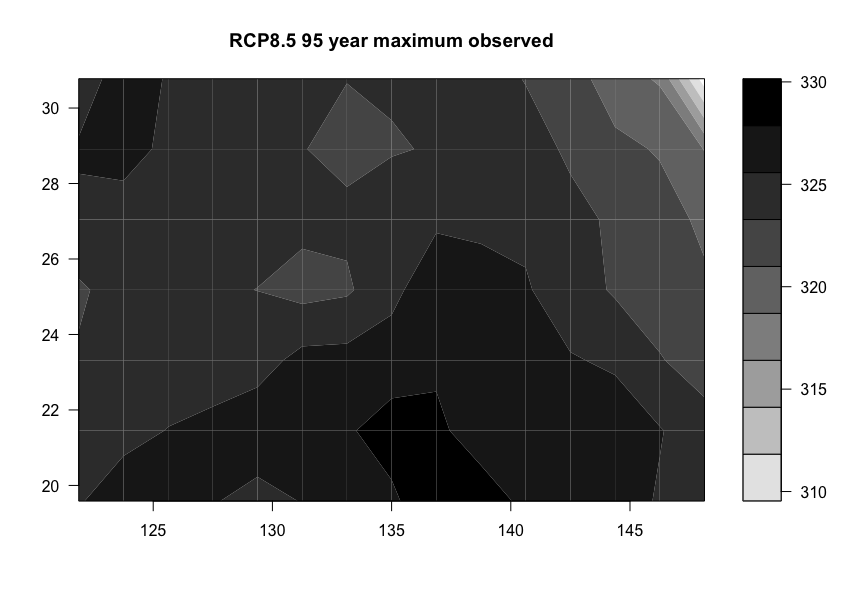}
 \caption{Predicted and observed 95-years return levels over Australia based on historical (top row), RCP4.5 (middle row) and RCP8.5 (bottom row) scenario data. Columns denote predictions based on $B^2=15\times 15$ (left) and $B^2=30\times 30$ (middle) histograms and interpolated observed maxima (right).
 }
\label{fig:returns}
\end{figure}

Figure \ref{fig:returns} illustrates expected and observed (columns) $95$-year return levels for each dataset (rows) for $B=15, 30$. Higher expected (and observed) returns for the RCP4.5 and RCP8.5 scenarios compared to the historical setting are apparent.

Because extrapolation into and beyond the tails of observed data is sensitive to a model's parameter estimates, there are some differences in the return levels for the different values of $B$. This suggests that, for applications in spatial extremes at least, higher resolution histograms may be required, depending on the nature of inference required.

%
%
%
\section{Discussion}
\label{sec:7}

In this article we have introduced a novel method for constructing composite likelihood functions for histogram-valued random variables. Working with random histograms as summaries of large datasets allows for computational efficiencies, as the histograms can efficiently represent large amounts of data in a concise form. The benefit of working with composite likelihoods in this setting is that the inefficiencies of working with histograms for higher dimensional data can largely be avoided.

Our theoretical results show that if the bins in each random histogram are the same, then the symbolic composite MLE can be computed exactly by combining the data into a single histogram (by summing the totals in each bin). As the majority of the computational time for an analysis is spent in optimising the likelihood, this is a particularly useful result that can lead to fast inference. The precision of the {\color{black}composite MLE}, however, depends on the number of histograms: the more there are (assuming equal numbers of datapoints in each histogram), the lower the estimated variance of the composite MLE. This will either present hard limits on the possible level of inferential precision (if pre-made histograms are presented directly to the analyst), or allow a trade-off of precision for computation to be made. As computation of the Godambe information matrix is trivial compared to estimation of the {\color{black}composite MLE}, if the full dataset is available, then a large number of histograms could be used for relatively low computational costs.

Our results have also shown the efficiency of standard composite likelihood techniques when the data are grouped into time blocks such that it is know which block any data point belongs to, but it is not known where the datapoint lies within each block.

We have not considered the question of how to best construct the random histograms.
This was considered in the present context by \cite{zhang+bs16} and \citet{beranger+ls18}.
Possible approaches could follow standard nonparametric arguments of histogram binwidth selection (e.g.~\cite{Scott1985}, \cite{Wand1997}) or more complex space-partitioning processes such as random trees, or alternatively be chosen to optimise pre-specified utility or loss functions. This is a current topic of active research. {\color{black}{In terms of determining a suitable number of bins $B$ to ensure a good approximation of the classical composite likelihood (and MLE), we used a naive approach Section \ref{sec:6}
in which $B$ was increased until there was only minor improvement in the inference. While this approach is practically viable (in that computational overheads can still be much lower than implementing the standard classical analysis),
ideally a method would be constructed to identify a priori the fixed number of bins required to optimise some criterion. This could be 
e.g.~a binwidth selection algorithm, or a loss-function based method etc.}}

One of our motivations for analysing the extremes of very large climate datasets is that, while exceptions exist, it is not uncommon for  statistical analysis to only occur independently at each spatial location, with very little work done to analyse the spatial dependence \citep{Huang2016}. In Section \ref{sec:6}, by fitting the Gaussian max-stable process to historical and future scenario Australian temperature data, we were able to explore changes in the spatial dependence structure that will accompany different levels of greenhouse gas emission levels in the coming years, and provide insight into the effects of these changes. It would be extremely challenging to perform these analyses, and others with even larger datasets, using standard techniques.

For the analysis of Australian temperature extremes, the data are presented as being located at the centre of a box within a grid. As such, the presented analysis
ignores the fact that the data actually arose from the entire box, and not just this point location.
One possible extension of the work in this article is to similarly treat the actual spatial locations of each datapoint within each grid box as unknown locations within a spatial histogram.

This would also allow datasets with extremely large numbers of locations ($K$) to be spatially aggregated into smaller datasets with spatial bins as the locations instead of pointwise coordinates, potentially drastically decreasing the computational cost and allowing the analysis of much higher dimensional data.

\section*{Acknowledgements}
We are grateful to Dr Markus Donat (Climate Change Research Centre, UNSW Sydney) for providing the climate dataset used in Section \ref{sec:6}. This research is supported by the UNSW Data Science Hub, and the Australian Research Council through the Australian Centre of Excellence for Mathematical and Statistical Frontiers  (ACEMS; CE140100049), and the Discovery Project scheme (FT170100079).

\bibliographystyle{chicago}

\end{document}